\documentclass[fleqn,usenatbib]{mnras}

\usepackage[T1]{fontenc}
\usepackage{ae,aecompl}

\usepackage{graphicx}	
\graphicspath{{figures/}}
\usepackage{amsmath,amssymb}	
\usepackage[usenames]{color}
\usepackage{mathrsfs}
\usepackage{array}
\usepackage{widetext}

\usepackage{hyperref}
\definecolor{linkblue}{rgb}{0,0,0.8}
\definecolor{linkgreen}{rgb}{0,0.5,0}
\definecolor{purple}{rgb}{0.78,0.18,0.77}
\hypersetup{pdfpagemode=UseNone, pdfstartview=FitH, linkcolor=linkblue, %
            citecolor=linkblue, urlcolor=linkblue}

\def\beq{\begin{equation}}
\def\eeq{\end{equation}}

\def\la{\langle}
\def\ra{\rangle}
\newcommand{\lp}{\left(}
\newcommand{\rp}{\right)}
\newcommand{\lb}{\left[}
\newcommand{\rb}{\right]}

\newcommand\numberthis{\addtocounter{equation}{1}\tag{\theequation}}

\def\d{\partial}

\newcommand{\vk}{\boldsymbol{k}}
\newcommand{\vq}{\boldsymbol{q}}

\newcommand{\vx}{\boldsymbol{x}}

\newcommand{\kl}{k_{\rm L}}
\newcommand{\ks}{k_{\rm S}}

\newcommand{\dirac}{\delta_{\rm D}}

\newcommand{\invMpc}{h\, {\rm Mpc}^{-1}\,}
\newcommand{\hinvMsun}{h^{-1}M_\odot}

\title[Baryonic effects on the matter bispectrum]{Baryonic effects on the matter bispectrum}

\author[Foreman, Coulton, Villaescusa-Navarro, and Barreira]{%
Simon Foreman$^{1,2,3}$,
William Coulton$^{4}$,
Francisco Villaescusa-Navarro$^{5,6}$, 
\newauthor
and Alexandre Barreira$^{7}$
\\
$^{1}$Perimeter Institute for Theoretical Physics, 31 Caroline Street North, Waterloo, ON N2L 2Y5,
Canada \\
$^{2}$Dominion Radio Astrophysical Observatory, Herzberg Astronomy \& Astrophysics Research Centre, National Research Council Canada, \\ P.O.~Box 248, Penticton, BC V2A 6J9, Canada \\
$^{3}$Canadian Institute for Theoretical Astrophysics, University of Toronto, 60 St. George Street, Toronto, ON M5S 3H8, Canada \\
$^{4}$Institute of Astronomy and Kavli Institute for Cosmology Cambridge, Madingley Road, Cambridge, CB3 0HA, UK \\
$^{5}$Department of Astrophysical Sciences, Princeton University, Peyton Hall, Princeton NJ 08544-0010, USA \\
$^{6}$Center for Computational Astrophysics, Flatiron Institute, 162 5th Avenue, New York, NY 10010, USA \\
$^{7}$Max-Planck-Institut f\"{u}r Astrophysik, Karl-Schwarzschild-Str. 1, 85741 Garching, Germany
}

\pubyear{2020}

\begin{document}
\label{firstpage}
\pagerange{\pageref{firstpage}--\pageref{lastpage}}
\maketitle

%--------------------------------------------------------------------------------------
% ABSTRACT
%--------------------------------------------------------------------------------------
\begin{abstract}
The large-scale clustering of matter is impacted by baryonic physics, particularly AGN feedback. Modelling or mitigating this impact will be essential for making full use of upcoming measurements of cosmic shear and other large-scale structure probes. We study baryonic effects on the matter bispectrum, using measurements from a selection of state-of-the-art hydrodynamical simulations: IllustrisTNG, Illustris, EAGLE, and BAHAMAS.
We identify a low-redshift enhancement of the bispectrum, peaking at $k\sim 3h\,{\rm Mpc}^{-1}$, that is present in several simulations, and discuss how it can be associated to the evolving nature of AGN feedback at late times. This enhancement does not appear in the matter power spectrum, and therefore represents a new source of degeneracy breaking between two- and three-point statistics. In addition, we provide physical interpretations for other aspects of these measurements, and make initial comparisons to predictions from perturbation theory, empirical fitting formulas, and the response function formalism. We publicly release our measurements (including estimates of their uncertainty due to sample variance) and bispectrum measurement code as resources for the community.
\end{abstract}

% Select between one and six entries from the list of approved keywords.
% Don't make up new ones.
 \begin{keywords}
cosmology: theory -- large-scale structure of Universe -- methods: numerical -- galaxies: haloes
 \end{keywords}

%%%%%%%%%%%%%%%%% BODY OF PAPER %%%%%%%%%%%%%%%%%%

%--------------------------------------------------------------------------------------
% INTRO
%--------------------------------------------------------------------------------------
\section{Introduction}

Now that a variety of observational probes have converged on a standard cosmological model~\citep{Akrami:2018vks,Abbott:2017wau}, the field of cosmology is shifting its focus to detecting subtle effects that could provide clues about the nature of dark matter, dark energy, and the early Universe. Examples of effects that could be seen in galaxy surveys include scale-dependent galaxy bias induced on large scales by primordial non-Gaussianity~\citep{Dalal:2007cu}, suppressed clustering on small scales caused by massive neutrinos~\citep{Lesgourgues:2012uu} or certain dark matter candidates (e.g.~\citealt{Hui:2016ltb}), deviations from general relativity in the growth rate of structures~\citep{Huterer:2013xky}, or inflationary features in the spectrum of primordial perturbations that survive until late times~\citep{Slosar:2019gvt,Beutler:2019ojk}. Robustly constraining any of these effects will require exquisite control over the relevant theoretical predictions, incorporating small systematic effects that may have been irrelevant in previous analyses but will become significant as the precision of our data increases.

Many analyses of large-scale structure focus on two-point statistics of the matter density, for which the fundamental theoretical quantity is the matter power spectrum. To capitalize on the constraining power of future surveys of lensed galaxy shapes (``cosmic shear"), our predictions for the matter power spectrum must be accurate to 1\% or better up to at least $k\sim 5\invMpc$~\citep{Huterer:2004tr,Hearin:2011bp}, and a host of other observational contaminants, such as uncertainties in the telescope point-spread function or photometric redshifts, must also be well-controlled~\citep{Mandelbaum:2017jpr}. To make effective use of galaxy clustering, one requires a through understanding of galaxy bias~\citep{Desjacques:2016bnm}, redshift-space distortions~\citep{2011RSPTA.369.5058P}, and selection effects (e.g.~\citealt{Obuljen:2019ukz}).

Beyond these complications, one must account for the effects of baryons on large scale clustering. The matter power spectrum is typically modelled (either analytically or in N-body simulations) under the assumption that only gravitational forces are active at the relevant scales, but in reality, the heating and cooling of gas, the formation of stars and galaxies, and energy output from supernovae and active galactic nuclei (AGN) can all impact the evolution of the matter density at cosmological distances~\citep{White:2004kv,Zhan:2004wq,Jing:2005gm,Rudd:2007zx,2011MNRAS.415.3649V}. In particular, \cite{2011MNRAS.415.3649V} found that feedback from AGN can significantly alter the matter power spectrum at wavenumbers as low as~$0.3\invMpc$, prompting investigations into strategies to mitigate this impact in data analysis. See \cite{vanDaalen:2019pst} for a recent update of \cite{2011MNRAS.415.3649V}, and \cite{Chisari:2019tus} for a summary of the state of the art for two-point cosmic shear statistics. In addition, an understanding of the baryon distribution will be important for making full use of upcoming measurements of kSZ tomography \citep{Battaglia:2016xbi,Smith:2018bpn} and CMB lensing \citep{Chung:2019bsk}.

However, two-point statistics do not exhaust the information content of the matter density field, due to the non-Gaussianity of the density field at later times. A natural way to access additional statistical information is to examine higher-point statistics, beginning with the three-point correlation function, known as the bispectrum in Fourier space. The raw signal to noise in the matter bispectrum is a few times smaller than in the power spectrum, although it increases slightly faster for the bispectrum than for the power spectrum as a function of the maximum wavenumber considered~\citep{Kayo:2012nm,Chan:2016ehg,Rizzato:2018whp,Barreira:2019icq}.
Despite this fact, the bispectrum responds to a variety of effects distinctly from the power spectrum, enabling analyses that include bispectrum measurements to substantially improve constraints on standard cosmological parameters~\citep{Kayo:2013aha,Hahn:2019zob}, neutrino mass~\citep{Coulton:2018ebd,Chudaykin:2019ock,Hahn:2019zob}, galaxy bias parameters~\citep{Yankelevich:2018uaz}, and primordial non-Gaussianity~\citep{Karagiannis:2018jdt}. Realizing these improvements will require an understanding of the impact of different systematics, including baryonic effects, on the bispectrum, and this is a primary motivator for this paper.

Furthermore, baryonic effects on large-scale structure are of course intrinsically interesting, with the potential to shed new light on a variety of open questions in astrophysics, such as the details of AGN feedback and galaxy formation. Thus, it is natural to ask whether the bispectrum could supply additional information to help us answer these questions, as investigated for the power spectrum in~\cite{Harnois-Deraps:2014sva} and~\cite{Foreman:2016jzy}. 

In this paper, we seek to explore these questions by examining the matter bispectrum in several recent hydrodynamical simulations, each with different numerical methods and implementations for the behavior of baryons in cosmological volumes. By comparing measurements from IllustrisTNG\footnote{\url{http://www.tng-project.org}}, Illustris\footnote{\url{http://www.illustris-project.org}}, BAHAMAS\footnote{\url{http://www.astro.ljmu.ac.uk/~igm/BAHAMAS/}}, and EAGLE\footnote{\url{http://icc.dur.ac.uk/Eagle/}}, we quantify the impact of baryons on the matter bispectrum as compared to the power spectrum, discuss physical interpretations of this impact where possible, and take first steps toward comparing these measurements to different modelling approaches. We publicly release our power spectrum and bispectrum measurements\footnote{\label{foot:meas}These measurements are located at \url{https://github.com/sjforeman/hydro_bispectrum}.}, along with a new code for measuring bispectra from numerical simulations\footnote{\label{foot:code}This code is located at \url{https://github.com/sjforeman/bskit}, where we also release the scripts we used for other simulation processing and measurements in this work.}, in the hope that the community will use these measurements to develop modelling and analysis approaches (or test existing approaches) for the bispectrum that account for baryonic effects, just as power spectra from \cite{2011MNRAS.415.3649V}\footnote{\url{http://vd11.strw.leidenuniv.nl}} have been used in the past.

This paper is organized as follows. In Sec.~\ref{sec:methods}, we describe our procedures for measuring power spectra and bispectra, and estimating the associated sample variance uncertainties. In Sec.~\ref{sec:measurements}, we present our measurements from IllustrisTNG (\ref{sec:illustrisTNG}), Illustris (\ref{sec:ill}), and BAHAMAS and EAGLE (\ref{sec:baheagle}), and discuss the physical origins of various aspects of their dependences on wavenumber. We focus on IllustrisTNG in particular, where we find a distinctive feature in the matter bispectrum that we interpret using the halo model. In Sec.~\ref{sec:largescales}, we quantify the impact of baryons on the matter bispectrum on quasi-linear scales in the different simulations, and examine how this impact changes with wavenumber. In Sec.~\ref{sec:modellingApproaches}, we compare our measurements to two approaches for modelling the bispectrum: the fitting functions from~\cite{Scoccimarro:2000ee} and~\cite{GilMarin:2011ik} (Sec.~\ref{sec:scgm-fitfunctions}), and the response function formalism (Sec.~\ref{sec:responses}). We conclude and indicate possible future directions in Sec.~\ref{sec:conclusions}.

%--------------------------------------------------------------------------------------
% SIMULATION POWER SPECTRA AND BISPECTRA
%--------------------------------------------------------------------------------------
\section{Simulation power spectra and bispectra}
\label{sec:methods}

\subsection{Measurement procedure}
\label{sec:procedure}

We make our measurements on simulation outputs consisting of the positions and masses of various tracer particles at a given time. The simulations track the evolution of dark matter, gas, stars, and black holes, with a separate type of tracer particle associated with each. For each constant-time snapshot, we assign mass to a regular grid using a cloud-in-cell (CIC) interpolation scheme, and compute the mass overdensity $\delta_M(\vx)$ from the gridded mass $M(\vx)$ via $\delta_M(\vx) \equiv (M(\vx)-\bar{M})/\bar{M}$, where $\bar{M}$ is the average mass per cell. We do this separately for dark matter and ``baryon" particles (defined here as gas, star, and black hole particles), as well as for the total mass. Note that the total matter overdensity can be formed from the dark matter and baryon overdensities via
\beq
\delta_{\rm m} = f_{\rm c} \delta_{\rm c} + f_{\rm b} \delta_{\rm b}\ ,
\label{eq:delta-matter-from-components}
\eeq
where $f_\alpha \equiv \Omega_\alpha/(\Omega_{\rm b}+\Omega_{\rm c})$.

After Fourier transforming each overdensity field, we deconvolve the window function corresponding to the CIC mass assignment (e.g.~\citealt{Jing:2004fq}). We then wish to measure power spectra $P(k)$ and bispectra $B(k_1,k_2,k_3)$, defined by (assuming $\delta$ has isotropic statistics)
\begin{align}
\left\la \delta(\vk) \delta^*(\vk') \right\ra &= (2\pi)^3 \dirac(\vk-\vk') P(k)\ , \\
\left\la \delta(\vk_1) \delta(\vk_2) \delta(\vk_3) \right\ra
	&= (2\pi)^3 \dirac(\vk_1+\vk_2+\vk_3) B(k_1,k_2,k_3)\ .
\end{align}
We measure the power spectrum within a given wavenumber bin (which we take to be a spherical shell with some width $\Delta k$) by averaging all values of $|\delta(\vk)|^2$ within that bin%
\footnote{The Fourier modes contained in a simulation are discrete, so the expressions we use in Sec.~\ref{sec:procedure} must be replaced by appropriate discrete approximations.}:
\beq
\hat{P}(k) = \frac{1}{V_{\rm bin}}\int_{\rm bin} d^3\vq\,  |\delta(\vq)|^2\ ,
\label{eq:pest}
\eeq
where $V_{\rm bin} \equiv \int_{\rm bin} d^3\vq$ is the bin volume.

A brute-force measurement of the bispectrum would proceed in a similar way: for wavenumber bins for $\vk_1$, $\vk_2$, and $\vk_3$, we would average over all values of $\delta(\vk_1)\delta(\vk_2)\delta(\vk_3)$ such that each $\vk_i$ was within its respective bin and the triangle condition $\vk_1+\vk_2+\vk_3$ was satisfied:
\begin{align*}
\hat{B}(k_1,k_2,k_3) &=  \frac{1}{V_\Delta}
	\int_{{\rm bin}\,1} d^3\vq_1 \int_{{\rm bin}\,2} d^3\vq_2 \int_{{\rm bin}\,3} d^3\vq_3 \\
&\quad \times \delta(\vq_1)\delta(\vq_2)\delta(\vq_3) \dirac(\vq_1+\vq_2+\vq_3)\ ,
	\numberthis
	\label{eq:best-brute}
\end{align*}
where
\beq
V_\Delta \equiv 
	\int_{{\rm bin}\,1} d^3\vq_1 \int_{{\rm bin}\,2} d^3\vq_2 \int_{{\rm bin}\,3} d^3\vq_3
	\, \dirac(\vq_1+\vq_2+\vq_3)\ .
	\label{eq:vDelta}
\eeq
However, we instead use an algorithm that exploits Fourier transforms to achieve greater computational efficiency than a manual iteration over all triangle configurations (e.g.~\citealt{Scoccimarro:2000sn,Sefusatti:2015aex}; our description follows \citealt{Tomlinson:2019bjx}). This algorithm is derived by expanding the Dirac delta function in plane waves,
\beq
\dirac(\vq_1+\vq_2+\vq_3)  = \int \frac{d^3\vx}{(2\pi)^3} e^{-i(\vq_1+\vq_2+\vq_3)\cdot\vx}\ ,
\eeq
inserting this into Eq.~\eqref{eq:best-brute}, and grouping factors that depend on each $\vq_i$. The result is
\beq
\hat{B}(k_1,k_2,k_3) = \frac{1}{V_\Delta} \int \frac{d^3\vx}{(2\pi)^3} I_{k_1}(\vx) I_{k_2}(\vx) I_{k_3}(\vx)\ ,
\label{eq:best-fast}
\eeq
where
\beq
I_{k_i}(\vx) \equiv \int d^3\vq\, W_i(\vq) e^{-i\vq\cdot\vx} \delta(\vq)
\label{eq:Iki}
\eeq
and $W_i(\vq)$ is unity if $\vq$ is within the bin centered on~$k_i$ and zero otherwise. In this way, $\hat{B}$ is reduced to a sequence of Fourier transforms and multiplications, which are easily parallelizable. We have written a Python module (see footnote~\ref{foot:code})
that implements this algorithm using routines from the public \texttt{nbodykit} package\footnote{\url{https://github.com/bccp/nbodykit}} \citep{Hand:2017pqn}. Other manipulations for this paper, including mass assignment and power spectrum estimation, have been performed in \texttt{nbodykit} and \texttt{Pylians}\footnote{\url{https://github.com/franciscovillaescusa/Pylians}}. 

Finally, the measured power spectra and bispectra will have extra shot-noise power arising from the discreteness of the tracer particles used to represent the density fields. Under the assumption that the tracer particles Poisson-sample the corresponding continuous fields, the contributions from white shot noise can be derived following the methods in~\cite{Feldman:1993ky} or~\cite{Chan:2012jx}, and are given by
\begin{align}
P_{\rm shot}(k) &= \frac{1}{\bar{n}} \frac{\left\la m^2 \right\ra}{\left\la m \right\ra^2}\ , \\
B_{\rm shot}(k_1,k_2,k_3) &= \frac{1}{\bar{n}^2} \frac{\left\la m^3 \right\ra}{\left\la m \right\ra^3}
	+ \frac{1}{\bar{n}} \frac{\left\la m^2 \right\ra}{\left\la m \right\ra^2}
	\lb P(k_1) + \text{2 perms} \rb\ ,
\end{align}
where $\la m \ra$, $\la m^2 \ra$, and $\la m^3 \ra$ are averages over the masses of all tracer particles. For the lowest-resolution simulations we consider (BAHAMAS), these shot noise spectra are no more than 6\% of the measured power spectra and 1\% of the measured bispectra, respectively, at the highest wavenumbers and redshifts we use. In the measurement ratios we focus on, the induced relative error is of the same order, which is small enough that it will not affect any of our conclusions. While predictions for shot noise spectra are sometimes subtracted from simulation measurements, this procedure becomes dangerous at higher redshifts, where the Poisson-sampling hypothesis is untrue at larger scales (e.g.~\citealt{Sirko:2005uz}). Therefore, in this work, we do not implement any shot noise subtraction.

\subsection{Grid resolution and binning scheme}

For each simulation, we assign mass to a grid with $2048^3$ cells, and then downsample the grid to $1024^3$ cells before making our measurements. This downsampling is accomplished in Fourier space by low-pass filtering the density grid. This does not introduce any error into the power spectrum estimated from the lower-resolution grid, since the lower-$k$ Fourier modes are untouched by this procedure and we use the binned estimator from Eq.~\eqref{eq:pest}.

Other errors could arise from particle discreteness or aliasing on the density grid. To ensure that particle discreteness errors are at the sub-percent level, we take the maximum reliable wavenumber of our power spectrum measurements, $k_{\rm max}^P$, to satisfy $k_{\rm max}^P < k_{\rm Ny}^{\rm part,DM}/2$, where $k_{\rm Ny}^{\rm part,DM} = \pi N_{\rm part,DM}^{1/3}/L_{\rm box}$ is the dark matter particle Nyquist frequency~\citep{Heitmann:2008eq,Schneider:2015yka}. To suppress aliasing to the same level, we also take $k_{\rm max}^P < k_{\rm Ny}^{2048}/2$, where $k_{\rm Ny}^{2048} = 2048\pi/L_{\rm box}$ is the Nyquist frequency of the original mass assignment grid~\citep{Sefusatti:2015aex}\footnote{Note that it is the resolution of the original grid, rather than the downsampled grid, that determines the Nyquist frequency for the purpose of estimating errors due to aliasing: aliasing is introduced when mass is assigned to the original grid, and the Fourier modes on the downsampled grid are identical to those from the original grid.}.

We take similar criteria for the maximum wavenumber~$k_{\rm max}^B$ at which the measured bispectrum is reliable, accounting for the fact that aliasing is introduced at a frequency slightly smaller than the Nyquist frequency of the subsampled grid used for the measurements, due to the Fourier transforms and multiplications involved in evaluating Eqs.~\eqref{eq:best-fast} and~\eqref{eq:Iki} \citep{Sefusatti:2015aex,Hung:2019ygc}. Thus, we take $k_{\rm max}^B < k_{\rm Ny}^{1024}/2$, in addition to $k_{\rm max}^B < k_{\rm Ny}^{\rm part,DM}/2$ (which is automatically satisfied by the first condition). 
See Appendix~\ref{app:aliasing} for a more detailed discussion of aliasing in bispectrum measurements, including a brief comparison of the claims of~\cite{Sefusatti:2015aex} and~\cite{Hung:2019ygc}.
Table~\ref{tab:kmax} summarizes the resulting $k_{\rm max}^P$ and $k_{\rm max}^B$ values for each simulation we consider. 

\begin{table}
	\centering
	\caption{Simulations used in this work. The two rightmost columns indicate the maximum wavenumbers up to which our power spectrum and bispectrum measurements are trustworthy, following the criteria described in the main text.}
	\label{tab:kmax}
	\begin{tabular}{lccccc} 
		\hline
		Simulation & $L_{\rm box}$ & $N_{\rm part,DM}^{1/3}$ 
			& $k_{\rm f}$ & $k_{\rm max}^P$ & $k_{\rm max}^B$ \\
		& $[h^{-1}\,{\rm Mpc}]$ & & \multicolumn{3}{c}{$[\invMpc]$} \\
		\hline
		TNG300 & 205 & 2500 & 0.031 & 15.7 & 7.8 \\
		TNG100 & 75 & 1820 & 0.084 & 38.1 & 21.5 \\
		Illustris & 75 & 1820 & 0.084 & 38.1 & 21.5 \\
		EAGLE & 67.7 & 1504 & 0.093 & 34.9 & 23.8 \\
		BAHAMAS & 400 & 1024 & 0.016 & 4.0 & 4.0 \\
		\hline
	\end{tabular}
\end{table}

For our power spectrum measurements, we use wavenumber bins with width equal to the fundamental wavenumber of each simulation box, $k_{\rm f} \equiv 2\pi/L_{\rm box}$. For the bispectrum, we use a bin width of $k_{\rm f}$ for each triangle side with $k_i<40k_{\rm f}$, and width $6k_{\rm f}$ for $40k_{\rm f}<k_i<k_{\rm max}^B$. This results in 76 or 77 linear $k$ bins, and between 21,000 and 22,000 bins of unique $(k_1,k_2,k_3)$ combinations, per simulation snapshot, depending on the simulation.
For each bispectrum bin, we also quote the average $k_i$ values over the bin, computed by inserting $q_i$ into the integral in Eq.~\eqref{eq:vDelta} and dividing the result by $V_\Delta$. See footnote~\ref{foot:meas} for the public release of these measurements.

\subsection{Estimates of sample variance}
\label{sec:errorbars}

Power spectra and bispectra measured in hydro and ``dark matter only" (i.e.\ gravity-only N-body, hereafter referred to as DMO) simulations will be subject to sample variance: the configuration of late-time structures in a given simulation run will depend on the realization of the initial conditions, which will vary between different runs. Some of this variance will cancel when we compute the ratio of measurements made on a hydro-DMO simulation pair, since each pair has the same initial conditions, but the number and properties of the higher-mass halos most strongly affected by feedback will differ from pair to pair, and therefore the impact of baryons on the measured statistics will also differ. Ideally, this variance could be estimated using an ensemble of simulations with different initial conditions but identical subgrid models, resolutions, and cosmological parameters. However, due to the computational cost of running the simulations we have used in this work, it would be too expensive to create such an ensemble solely for the purpose of estimating the sample variance.

Therefore, we have opted instead to estimate the sample variance using measurements performed in sub-boxes of each simulation. Specifically, we have measured the power spectrum and bispectrum in each of 8 sub-boxes, evaluated the hydro/DMO ratio, and examined the spread in these measurements. To quantify the spread, we compute the 68\% range (centered on the median) of the measurements\footnote{We compute this as the difference between the 16th and 84th percentiles of the set of measurements for the 8 sub-boxes, with each percentile found by linearly interpolating between the two closest measured values. This is implemented, for example, in the \texttt{scipy} function \texttt{scipy.stats.iqr(vals,rng=[16,84])} where \texttt{vals} is the list of values.}, divide by 2, and further scale this by a factor of~$1/\sqrt{8}$ (the square root of the ratio of sub-box and full volumes). For a Gaussian distribution, this would be equal to an estimate of the standard deviation, but our measured ratios have heavier tails than a Gaussian distribution, leading to occasional outlier points that massively increase the standard deviation itself (see Appendix~\ref{app:spikes} for more details). We choose instead to define errorbars that reflect the typical distribution of measurements around the true value, as given by the central 68\% range, since this definition is not influenced as much by outliers.

If one were to estimate the sample variance of the power spectra on their own (without taking any ratios) from sub-boxes, one would find a large extra contribution from so-called ``super-sample variance," arising from couplings between density modes within each subvolume and modes with longer wavelength~\citep{Hamilton:2005dx,Takada:2013bfn,Li:2014sga,Barreira:2017fjz}. However, for the ratio of power spectra with shared initial conditions, this super-sample contribution cancels to a very good approximation: intuitively, the same long modes modulate the local power in both the hydro and DMO simulations, so the power spectrum ratio is largely insensitive to this modulation. We justify this statement in more detail in Appendix~\ref{app:subboxes}, using both analytical descriptions of the super-sample variance and empirical comparisons of sub-box measurements with an ensemble of independent hydrodynamical simulations.\footnote{See also~\cite{Peters:2016ubo} and~\cite{Chisari:2018prw} for investigations of sample variance of baryon effects within sub-boxes of a single simulation, and~\cite{vanDaalen:2019pst} for comparisons between simulations with different initial conditions.} For the bispectrum, the super-sample variance is far subdominant to the intrinsic (non-Gaussian) variance on quasi-linear scales~\citep{Chan:2017fiv}, but on deeply nonlinear scales its impact is less clear~\citep{Kayo:2012nm,Rizzato:2018whp}, so we calibrate the variances measured from sub-boxes by comparing with the independent simulations mentioned above; see Appendix~\ref{app:subboxes} for details.

Of course, these estimates of the sample-variance uncertainty on power spectra and bispectra ratios will themselves be subject to sample variance, occasionally giving values for the uncertainty that are clearly over-estimated when compared to values at nearby wavenumbers. Since the level of sample variance should vary smoothly with wavenumber, when plotting the sample-variance ``errorbars" in this paper, we apply an additional smoothing procedure to the estimated uncertainties as a way to correct for these outliers. Specifically, we compare each estimated uncertainty, $\sigma_i$ (with $i$ indexing the relevant wavenumber bins), to the mean of the uncertainties in the two neighboring bins, $(\sigma_{i-1}+\sigma_{i+1})/2$. If $\sigma_i > 3(\sigma_{i-1}+\sigma_{i+1})/2$, we use the neighbor average instead of the directly estimated value of $\sigma_i$. This procedure generally smoothes away obvious outliers in the estimated errorbars. In our public release of the measurements, we do not apply any such smoothing of the reported errorbars, so the user is free to treat them as they desire.

%--------------------------------------------------------------------------------------
% MEASUREMENTS AND INTERPRETATIONS
%--------------------------------------------------------------------------------------
\section{Measurements and interpretations}
\label{sec:measurements}

In this section, we present and discuss our measurements from IllustrisTNG (\ref{sec:illustrisTNG}), Illustris (\ref{sec:ill}), and BAHAMAS and EAGLE (\ref{sec:baheagle}). We summarize these measurements in Sec.~\ref{sec:summary} (see Fig.~\ref{fig:summary}).

\subsection{IllustrisTNG} 
\label{sec:illustrisTNG}

We first present results from the IllustrisTNG simulations \citep{Pillepich:2017fcc,Springel:2017tpz,Nelson:2017cxy,Naiman:2018,Marinacci:2017wew,Nelson:2018uso}, a set of hydrodynamical simulations whose aim is to reproduce realistic properties of galaxies within cosmological volumes. They were run with the moving-mesh code AREPO~\citep{Springel:2009aa}, and build on numerical and modelling approaches originally implemented in the earlier Illustris simulations~\citep{Vogelsberger:2013eka}, incorporating numerous advances designed to mitigate the physical shortcomings that were identified in Illustris \citep{Nelson:2015dga}. These advances are described in \cite{Pillepich:2017jle} and \cite{2017MNRAS.465.3291W}, and include new implementations of galactic winds, stellar evolution, magnetohydrodynamics, and AGN feedback. In particular, AGN feedback occurs in either of two modes: ``thermal mode," in which the temperature of the nearby gas is continuously increased by an amount proportional to the mass accretion rate, or ``kinetic mode," in which nearby gas particles receive momentum kicks in proportion to both the mass accretion rate and the local gas density. For a given black hole, if the Bondi accretion rate $\dot{M}_{\rm Bondi}$ ($\propto M_{\rm BH}^2 \rho$, where $\rho$ is the density of the surrounding gas) is above some fraction of the Eddington rate $\dot{M}_{\rm Edd}$ ($\propto M_{\rm BH}$), thermal-mode feedback takes place, while if $\dot{M}_{\rm Bondi}$ is below this fraction, kinetic-mode feedback occurs. This fraction has its own dependence on the black hole mass; see~\cite{2017MNRAS.465.3291W} for further details.

The IllustrisTNG team has run simulations in three different box sizes, denoted by TNG300 ($L_{\rm box}=302.6\,{\rm Mpc}=205h^{-1}\,{\rm Mpc}$), TNG100 ($L_{\rm box}=110.7\,{\rm Mpc}=75h^{-1}\,{\rm Mpc}$), and TNG50 ($L_{\rm box}=51.7\,{\rm Mpc}=35h^{-1}\,{\rm Mpc}$; not yet public). TNG300 and TNG100 exist at three different mass resolutions; in this section, we present results from the highest-resolution run at each box size (denoted by the suffix ``-1"), corresponding to dark matter particle masses of $4\times 10^7\hinvMsun$ and $5\times 10^6\hinvMsun$ respectively. See Appendix~\ref{app:tng-resolution} for a comparison between the different resolutions at each box size. For each box size and resolution, a DMO  simulation with the same initial conditions has also been run. All runs use Planck 2015 cosmological parameters \citep{Ade:2015xua}.

\begin{figure*}
\includegraphics[width=0.985\textwidth]{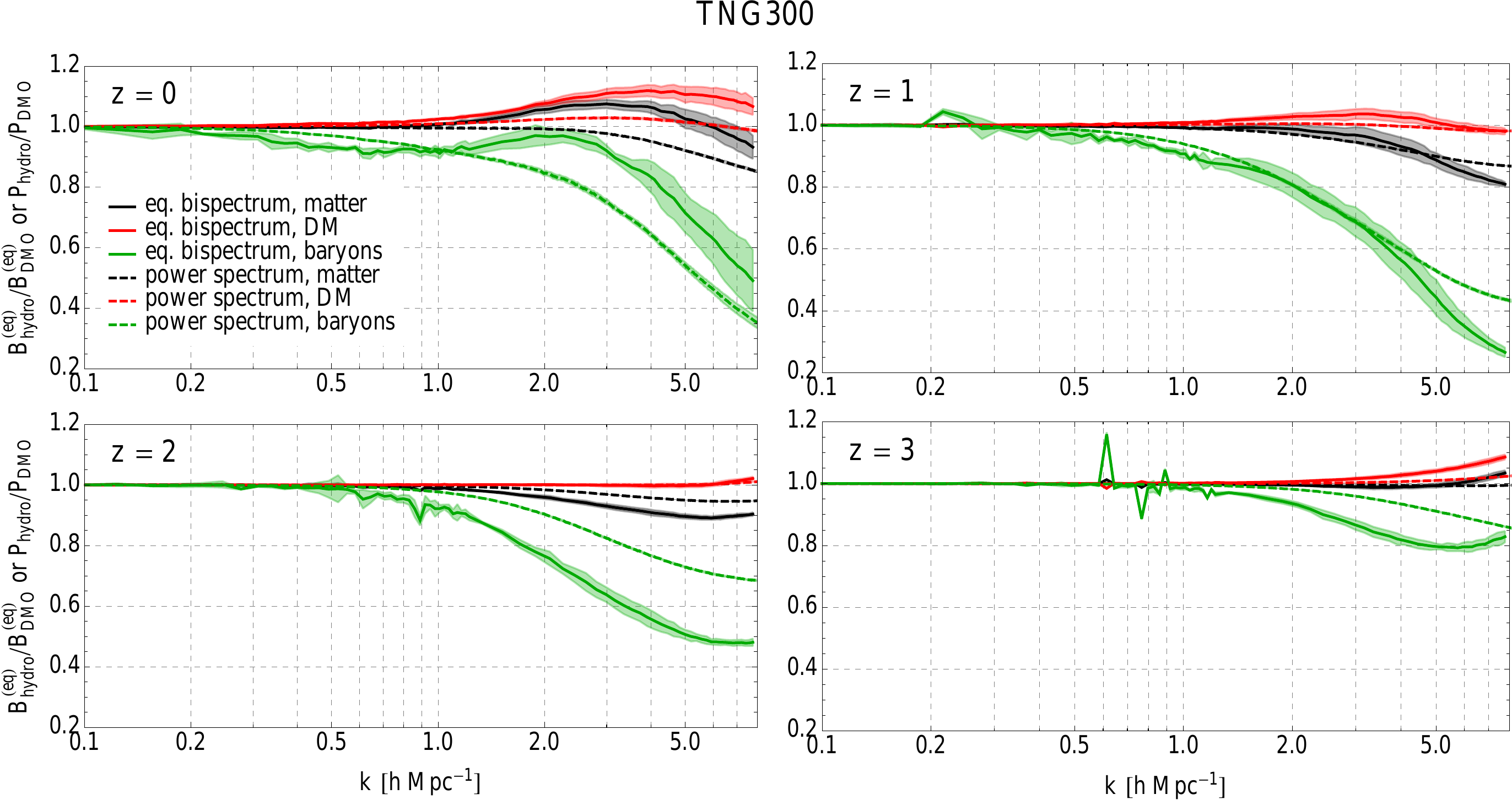}
\caption{
\label{fig:tng300-psbs}
Measured power spectra (dashed lines) and equilateral bispectra (solid lines) of matter (black), dark matter only (red), and baryons only (green) from TNG300-1, normalized to measurements from the DMO run. Estimates of the sample variance of each ratio are shown in the colored bands. The deepening suppressions of the ratios at later times are due to AGN feedback, while the enhancements peaking at $k\sim 3\invMpc$ can be attributed to backreaction of this feedback on the distribution of dark matter. Such an enhancement is seen in the matter bispectrum but not in the power spectrum at $z=0$, indicating that the former can provide qualitatively different information than the latter about baryonic processes in large-scale structure.
}    
\end{figure*}

\subsubsection{Simulation measurements}

\begin{figure*}
\includegraphics[width=0.985\textwidth]{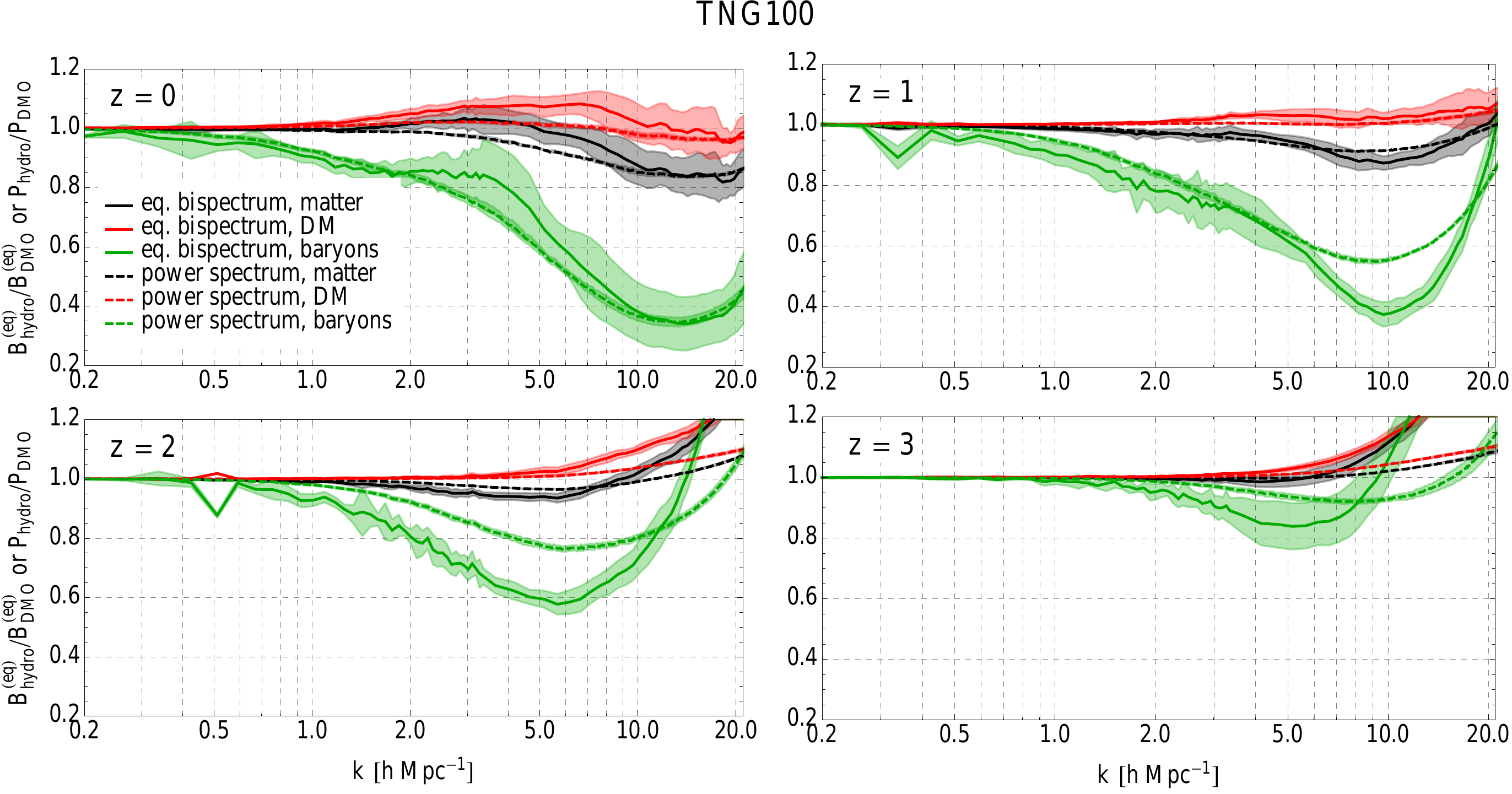}
\caption{
\label{fig:tng100-psbs}
Same as Fig.~\ref{fig:tng300-psbs}, but for measurements from TNG100-1. The same general trends are visible, with milder enhancements of the total matter and dark matter bispectra at late times, because TNG100-1 contains fewer of the high-mass halos that dominate the behavior of the spectra at these scales. Note the different $x$-axis range from Fig.~\ref{fig:tng300-psbs}.
}    
\end{figure*}

In Fig.~\ref{fig:tng300-psbs}, we show measurements from TNG300-1 of power spectra and equilateral bispectra of total matter, dark matter only, and baryons only, normalized to matter power spectra and bispectra measured from the DMO run. Fig.~\ref{fig:tng100-psbs} shows the same measurements from TNG100-1. Most of the large-scale sample variance present in each individual spectrum cancels out in the hydro/DMO ratio, resulting in smooth curves at low wavenumbers. An estimate of the  residual sample variance in the ratios is shown in the colored bands, and is constructed using the method from Sec.~\ref{sec:errorbars}.

We first discuss the matter power spectrum, previously presented for IllustrisTNG in~\cite{Springel:2017tpz}. A main visible feature in Figs.~\ref{fig:tng300-psbs}-\ref{fig:tng100-psbs} is a suppression of power in the hydro simulation as compared to the DMO run, generally deepening with decreasing redshift and increasing wavenumber (smaller comoving scale). At $z=3$, the suppression is visible only in the baryons (at TNG100-1 resolution the total matter even displays an enhancement), and is caused by the delayed collapse of gas into halos due to pressure in the baryonic material in the hydro simulations; note that this early-time suppression is also seen in simulations without AGN feedback~\citep{2011MNRAS.415.3649V,Chisari:2018prw}, indicating that such feedback typically plays a very minor role in large-scale structure evolution at $z\gtrsim 3$. In the TNG100-1 simulation, for $k \gtrsim 5-10\invMpc$, the baryon curves show an upturn attributable to the effects of adiabatic gas cooling. At lower redshifts, the increased suppression is attributable to feedback effects that heat up the gas within halos and distribute it over larger regions, smoothing out the clustering on these scales~\citep{2011MNRAS.415.3649V,Chisari:2019tus,vanDaalen:2019pst}, although, as we will discuss later, there is evidence that these effects do not monotonically increase with time. At $k\gtrsim 10\invMpc$, we see an upturn in the power spectrum compared to the DMO measurement, attributable to halo contraction caused by dense stellar (and, before feedback is strong, gas) mass in the halo centers. This upturn is more visible in power spectrum measurements that extend to smaller scales (e.g.~\citealt{2011MNRAS.415.3649V,vanDaalen:2019pst}), but is particularly susceptible to resolution effects in current simulations~\citep{vanDaalen:2019pst}.

Based on the physical picture of AGN feedback causing most of the suppression at $k\lesssim 10\invMpc$ and $z<3$, one might naively expect the matter bispectrum to show similar behavior: a monotonic drop in power that increases with wavenumber. In Figs.~\ref{fig:tng300-psbs}-\ref{fig:tng100-psbs}, we generally see this at $z\geq 1$. However, at lower redshift, we instead find an enhancement of the matter bispectrum above the DMO measurement, peaking at $k\sim 3\invMpc$ before entering a region of suppression at higher $k$. This enhancement is about equal to the sample-variance errorbars for TNG100, but is more clearly visible in TNG300. The appearance of a new physical scale, not immediately visible in the power spectrum ratio, indicates that the equilateral matter bispectrum can supply qualitatively distinct information about the feedback processes (or other baryonic effects) that alter large-scale structure in this regime. 

Previously,~\cite{Semboloni:2012yh} reached roughly the same conclusion: two- and three-point statistics---specifically, aperture mass statistics in weak lensing---can be used to break degeneracies between baryonic effects and background cosmology that are present in two-point statistics alone. However, they used a third-order statistic that effectively integrates the matter bispectrum over a wide wavenumber range and also imposes a specific redshift weighting, while we have measured the full bispectrum and can therefore examine its time- and scale-dependence. Furthermore, they relied on a fitting function that relates the matter bispectrum to the power spectrum and a fitted coefficient function. With our direct simulation measurements, we are able to assess the validity of this modelling approach; we will do so in Sec.~\ref{sec:scgm-fitfunctions}. 

To investigate the source of the matter bispectrum's scale-dependence, we now turn to the power spectrum and bispectrum of dark matter and baryons separately. The baryon statistics experience an upturn at the smallest scales (see, for example, in TNG100 at $z=2$), owing to galaxy formation in the centers of halos, while they are suppressed at larger scales, thanks to the action of AGN feedback. At $z=0$, the baryon power spectrum and bispectrum ratios become flatter than at earlier redshifts  between roughly $0.5\invMpc$ and a few $\invMpc$, with the latter even developing a local peak at $k\sim 2\invMpc$ in TNG300. Meanwhile, the dark matter statistics, just like the total-matter bispectrum, are enhanced at $k\gtrsim 1 \invMpc$ relative to their DMO counterparts. This enhancement of the dark matter power spectrum has been previously observed in several simulations, with the exception of those with the strongest AGN feedback~\citep{2011MNRAS.415.3649V,Hellwing:2016ucy,Springel:2017tpz,Chisari:2018prw,vanDaalen:2019pst}. We will re-examine some of these simulations, in light of our bispectrum measurements, later in this paper.

\begin{figure}
\includegraphics[width=\columnwidth]{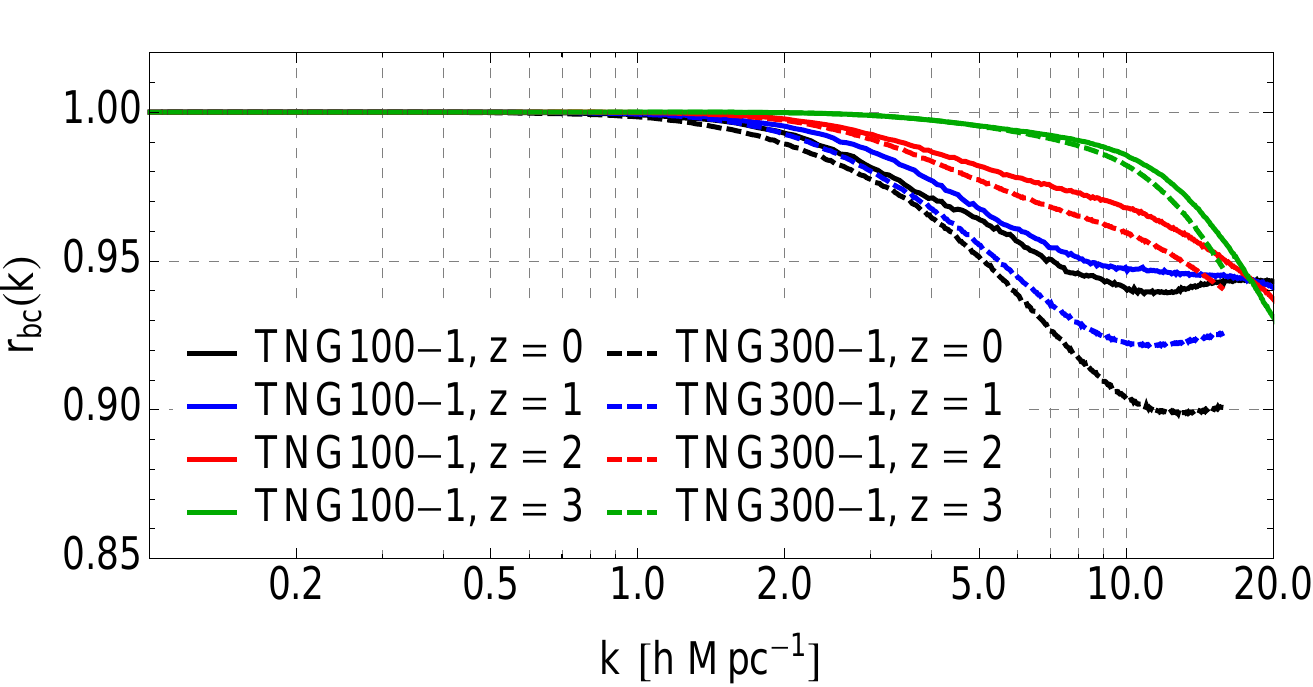}
\caption{\label{fig:rbc}
The baryon-dark matter cross-correlation coefficient, $r_{\rm bc}(k) \equiv P_{\rm bc}(k)/\lb P_{\rm bb}(k) P_{\rm cc}(k) \rb^{1/2}$, is greater than 90\% for $k\lesssim 10\invMpc$ in TNG100-1 (solid lines) and TNG300-1 (dashed lines). This implies that the phases of individual modes of $\delta_{\rm b}$ and $\delta_{\rm c}$ are almost identical at these wavenumbers, and therefore the cross power spectra and bispectra of $\delta_{\rm b}$ and $\delta_{\rm c}$ can be written in terms of the auto spectra to a very good approximation.
}    
\end{figure}

Before exploring the physical reason for this enhancement, we remind the reader that the statistics of the total matter overdensity are completely determined by the separate (auto and cross) statistics of the dark matter and baryons. Specifically, Eq.~\eqref{eq:delta-matter-from-components} implies that
\begin{align*}
\numberthis
\label{eq:pm-from-bc}
P_{\rm m}(k) &= f_{\rm c}^2 P_{\rm cc}(k) + f_{\rm b}^2 P_{\rm bb}(k) 
	+ 2f_{\rm c}f_{\rm b} P_{\rm bc}(k)\ , \\
B_{\rm m}(k_1,k_2,k_3) &= f_{\rm c}^3 B_{\rm ccc}(k_1,k_2,k_3)
	+ f_{\rm b}^3 B_{\rm bbb}(k_1,k_2,k_3) \\
&\quad + f_{\rm c}^2 f_{\rm b} \lb B_{\rm ccb}(k_1,k_2,k_3) + \text{2 perms} \rb \\
&\quad + f_{\rm c} f_{\rm b}^2 \lb B_{\rm cbb}(k_1,k_2,k_3) + \text{2 perms} \rb\ .
	\numberthis
	\label{eq:bm-from-bc}
\end{align*}
We can simplify these expressions by noting that the baryon-dark matter cross-correlation coefficient, $r_{\rm bc}(k) \equiv P_{\rm bc}(k)/\lb P_{\rm bb}(k) P_{\rm cc}(k) \rb^{1/2}$, is greater than 90\% for $k\lesssim 10\invMpc$ in TNG300-1 and TNG100-1 (see. Fig.~\ref{fig:rbc}). This implies that the phases of individual modes of $\delta_{\rm b}$ and $\delta_{\rm c}$ are almost identical at these wavenumbers, and therefore we can approximate the cross power spectra and equilateral cross bispectra by the geometric means of the autocorrelations:
\begin{align*}
P_{\rm bc}(k) &\approx \lb P_{\rm bb}(k)P_{\rm cc}(k) \rb^{1/2}\ , \\
B_{\rm bbc}^{\rm (eq)}(k) &\approx \lb B_{\rm bbb}^{\rm (eq)}(k)^2 B_{\rm ccc}^{\rm (eq)}(k) \rb^{1/3}\ , \\
B_{\rm bcc}^{\rm (eq)}(k) &\approx \lb B_{\rm bbb}^{\rm (eq)}(k) B_{\rm ccc}^{\rm (eq)}(k)^2 \rb^{1/3}\ .
	\numberthis
\end{align*}
Thus, we can approximate Eqs.~\eqref{eq:pm-from-bc} and~\eqref{eq:bm-from-bc} by
\begin{align*}
\numberthis
P_{\rm m}(k) &\approx \lb f_{\rm c} P_{\rm cc}(k)^{1/2} + f_{\rm b} P_{\rm bb}(k)^{1/2} \rb^2\ , \\
B_{\rm m}^{\rm (eq)}(k) &\approx 
	\lb f_{\rm c} B_{\rm ccc}(k)^{1/3} + f_{\rm b} B_{\rm bbb}(k)^{1/3} \rb^3\ ,
	\numberthis
\end{align*}
such that we can apply an understanding of the separate baryon and dark matter statistics to the total matter statistics by considering the appropriate weighted sum.

\subsubsection{Halo model interpretation}
\label{sec:halomodel}

Using the halo model (e.g.~\citealt{Cooray:2002dia}), the power spectrum and bispectrum differences between the hydro and DMO simulations can be attributed to modifications of the density profiles of group- and cluster-scale halos, caused by the evolving nature of AGN feedback at low redshift.

In the halo model, all matter is assumed to be contained within halos, allowing multi-point statistics to be modelled as a sum of ``$N$-halo terms" corresponding to correlations between points located in $N$ different halos. At sufficiently large wavenumbers, the power spectrum and bispectrum will be dominated by their respective one-halo terms, which are sensitive to the density profiles of halos within a certain mass range (see~\citealt{vanDaalen:2015msa} for an exploration of the relevant mass range as a function of $k$). These one-halo terms are given by
\begin{align}
\label{eq:p1h}
P_{\rm 1h}(k) &= \frac{1}{\bar{\rho}^2} \int dM\, M^2 W^2(k,M) \frac{dn}{dM}\ , \\
B_{\rm 1h}(k_1,k_2,k_3) &= \frac{1}{\bar{\rho}^3} \int dM\, M^3 
	\lb \prod_{i=1}^3 W(k_i,M) \rb \frac{dn}{dM}\ ,
	\label{eq:b1h}
\end{align}
where $dn/dM$ is the comoving number density of halos with a given mass, we have omitted the dependence on redshift, and $W(k,M)$ is the Fourier transform of the spherically-averaged halo profile for mass $M$:
\beq
W(k,M) \equiv \frac{1}{M} \int_0^{r_{\rm vir}} dr\, 4\pi r^2 \rho(r,M) \frac{\sin(kr)}{kr}\ .
\label{eq:wkm}
\eeq

These expressions can be applied to the total mass, and also to dark matter and baryons separately (with the caveat that the model has varying degrees of validity for different components, and is imperfect even for the total mass in DMO simulations \citep{Schmidt:2015gwz}). The different dependences of Eqs.~\eqref{eq:p1h} and~\eqref{eq:b1h} on $W(k,M)$ make it clear that the one-halo term for the bispectrum will respond more strongly than the power spectrum to mass-dependent changes in the halo profiles. In particular, the bispectrum has more sensitivity to the most massive halos, where AGN feedback plays an important role.

\begin{figure*}
\includegraphics[width=0.99\textwidth]{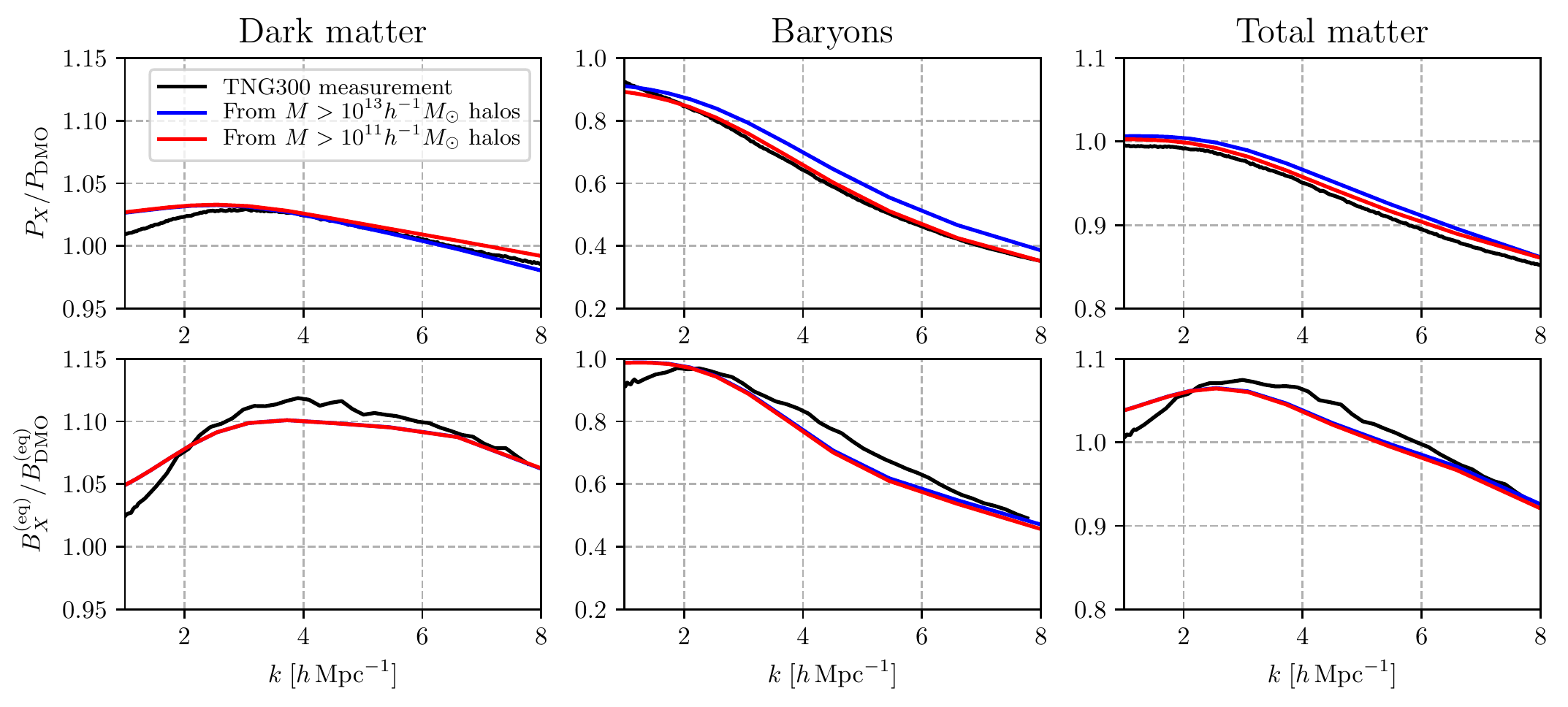}
\caption{
\label{fig:1halo-comparison}
We compare our measurements (black lines) for the ratios of power spectra (upper panels) and equilateral bispectra (lower panels) of dark matter, baryons, and total matter at $z=0$ to the corresponding one-halo contributions derived from the halo model, computed using the actual halo density profiles from TNG300-1 down to a minimum virial mass of $10^{13}\hinvMsun$ (blue lines) or $10^{11}\hinvMsun$ (red lines). The one-halo contributions provide a close enough match to the direct simulation measurements to justify interpreting the measured ratios in the context of the halo model. Also, the similarity of the red and blue lines confirms that halos more massive than  $10^{13}\hinvMsun$ mostly determine the behavior of the ratios over the wavenumber range of interest.
}    
\end{figure*}

To connect the halo model to our measured statistics, we have measured the baryon, dark matter, and total matter density profiles in all halos with virial mass greater than $10^{11} \hinvMsun$ in TNG300 and the associated DMO run. These profiles can then be used in discretized versions of Eqs.~\eqref{eq:p1h} and~\eqref{eq:b1h} (that simply sum over all measured profiles) to compute one-halo ``predictions" for the power spectrum and bispectrum ratios. In Fig.~\ref{fig:1halo-comparison}, we compare these computations to the measured ratios from TNG300 at $z=0$, finding sufficient agreement to justify interpreting the measured ratios using the density profiles of these halos. We find that halos with $M > 10^{13}\hinvMsun$ dominate the one-halo term at $k<10\invMpc$, in agreement with~\cite{vanDaalen:2015msa}. At wavenumbers smaller than those in Fig.~\ref{fig:1halo-comparison}, the one-halo term is no longer dominant, also in agreement with~\cite{vanDaalen:2015msa} for the power spectrum. 

\begin{figure*}
\includegraphics[width=0.99\textwidth]{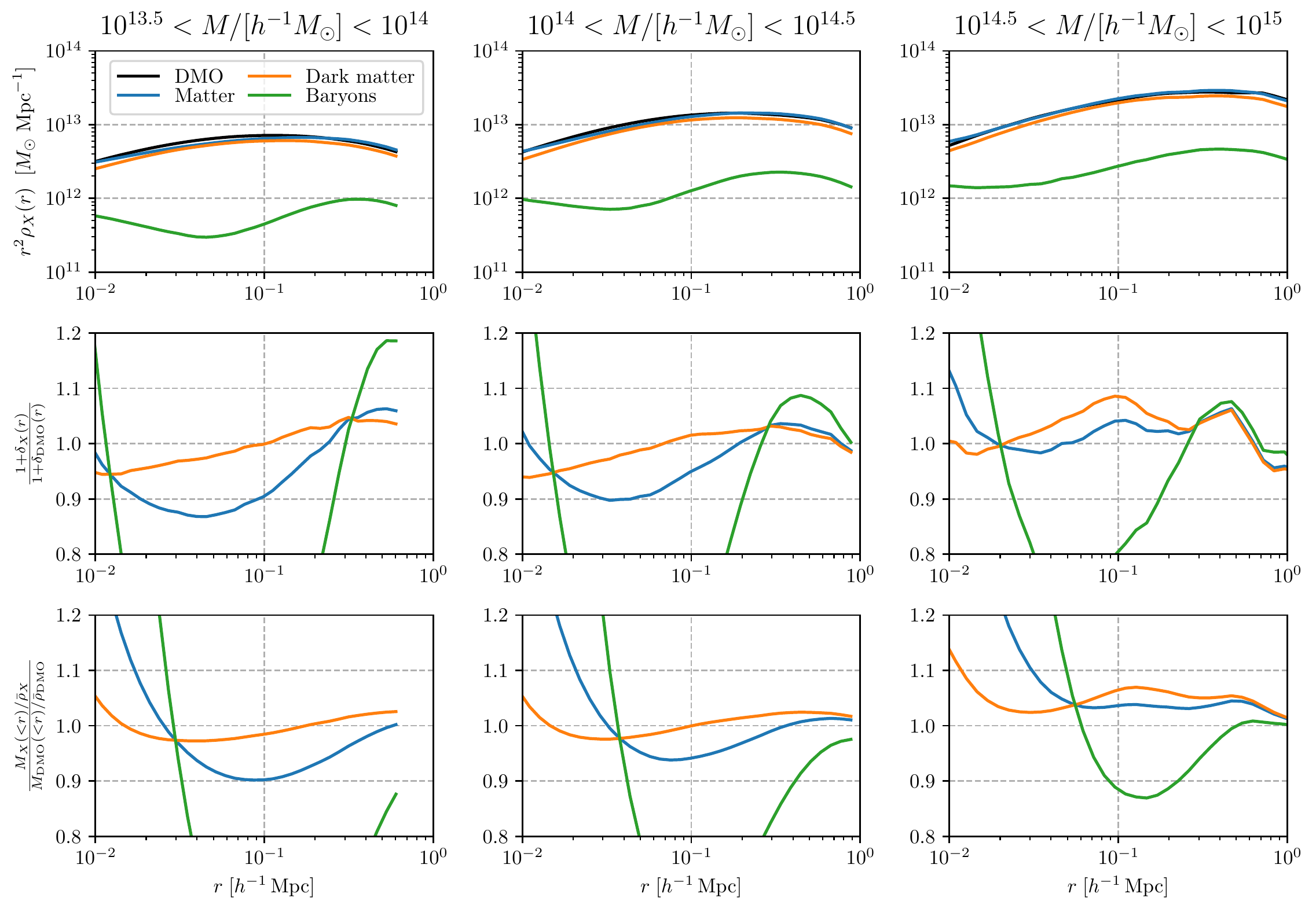}
\caption{
\label{fig:profiles}
{\em Upper panels:} Halo density profiles at $z=0$ from DMO run of TNG300-1 (black lines) and for total matter (blue lines), dark matter (orange lines), and baryons (green lines) from hydro run, averaged within the indicated bins in virial mass. The baryon profiles steepen at small radius due to concentrations of stellar mass where galaxies form, and peak at $r\sim 400h^{-1} {\rm kpc}$ due to weakened AGN feedback at late times (see main text for discussion). 
{\em Middle panels:} Ratio of density profiles in hydro and DMO runs, with each averaged profile normalized to the mean density for that particle type. The normalized baryon and dark matter profiles are enhanced relative to the DMO run for $r\gtrsim 300h^{-1}{\rm kpc}$, likely indicative of backreaction of the baryon distribution on the dark matter. The dark matter profiles are also enhanced at smaller radii, which will require more detailed halo-to-halo comparisons to fully explain.
{\em Lower panels:} Ratio of average mass enclosed within radius $r$ in hydro and DMO runs, again normalized to the mean density of each particle type. This enclosed mass is essentially what appears in the halo model as the normalized Fourier-transformed density profile [see Eq.~\eqref{eq:wkm}], and the radius at which the plotted ratio peaks is roughly the inverse of the wavenumber where the power spectrum or bispectrum ratios peak in our direct measurements.
}    
\end{figure*}

It is instructive to examine the halo profiles themselves, which we do in Fig.~\ref{fig:profiles}, averaged in three mass bins with $\Delta\log(M/[\hinvMsun])=0.5$. The panels in the top row show the profiles in the DMO run along with the total matter, dark matter, and baryon profiles in the hydro run. The DMO, total matter, and dark matter profiles broadly trace each other, while the baryon profiles steepen for $r\lesssim 30 h^{-1} {\rm kpc}$ due to the concentration of stellar mass there, and also rise to a second peak at $r\sim 400 h^{-1} {\rm kpc}$. This peak likely emerges because of a decrease in AGN feedback strength at late times. In IllustrisTNG, supermassive black holes with $M_{\rm BH}\gtrsim10^{8.5} \hinvMsun$ (which reside in the halos of interest here, with $M\gtrsim 10^{13}\hinvMsun$) experience rapid late-time decreases in the rates at which they accrete gas, with the accretion mode switching from thermal to kinetic~\citep{Weinberger:2017bbe}. This reduces the gas heating rate such that gas in the halo outskirts is no longer supported by feedback, and starts contracting towards the halo center. In turn, this backreacts on the gravitational potential and therefore on the dark matter at similar radii.

This interpretation is consistent with the low-redshift changes in the baryon curves in Fig.~\ref{fig:tng300-psbs}: from $z=1$ to $z=0$, the baryon power spectrum ratio increases in the range $1\invMpc\lesssim k \lesssim 4\invMpc$ and the bispectrum ratio increases over an even larger range. As opposed to $z>1$, where suppression that increases with time can be attributed to feedback expanding the baryonic material, the low-redshift reversal of this behavior at certain scales indicates that, on average, baryons are likely flowing inwards instead of being expelled.

The same phenomenon has been advanced to explain several features of the Horizon-AGN simulation~\citep{Dubois:2014lxa}. In this simulation, the dark matter power spectrum is enhanced at $z\lesssim 0.5$ at the same scale as in IllustrisTNG, and the gas fractions in massive halos are seen to grow at $z<1$~\citep{Chisari:2018prw}. Both features would arise if previously-expelled gas began re-accreting onto its original halos, and indeed, the largest black holes in Horizon-AGN have accretion rates that decline at late times~\citep{Volonteri:2016uhr}, implying that feedback likely decreases enough to allow this re-accretion to occur.

We can see the relative effects on each type of profile more clearly in the middle row of panels in Fig.~\ref{fig:profiles}, in which we plot the ratio of {\em over}density profiles, $[1+\delta_X(r)]/[1+\delta_{\rm DMO}(r)]$, since these profiles are what enter the one-halo term [recall Eqs.~\eqref{eq:p1h} to~\eqref{eq:wkm}]. The baryon overdensity profiles are enhanced relative to the DMO simulation for $r\gtrsim 300h^{-1}{\rm kpc}$ and suppressed for $r\lesssim 300h^{-1}{\rm kpc}$, and we also see an enhancement of the dark matter profiles at the same radii. However, in the dark matter profiles, this enhancement persists down to significantly smaller radii: for $10^{13.5}\hinvMsun < M < 10^{14.5}\hinvMsun$, we find enhancement for $r\gtrsim 100h^{-1}{\rm kpc}$, while for $M > 10^{14.5}\hinvMsun$, the enhancement stretches as low as $r\sim 20h^{-1}{\rm kpc}$, and also exhibits a second peak at $r\sim 100h^{-1}{\rm kpc}$.

Late-time gas re-accretion, combined with an understanding of how far out the dark matter was initially expelled, can likely account for some amount of dark matter density enhancement over the DMO case. However, a full explanation of the shapes of these profiles will require a detailed investigation of the redistribution of dark matter by baryonic effects within these halos, ideally cross-matching halos between hydro and DMO runs and tracking how the profiles evolve with time; this is beyond the scope of this work. Recent studies of halos in large hydrodynamical simulations~\citep{Velliscig:2014bza,Schaller:2014uwa,Peirani:2016qvp} have focused on more general properties such as mass functions or the presence of cored profiles. We advocate for increased attention to comparisons of DMO and hydro halos at $r\gtrsim 100h^{-1}{\rm kpc}$, since our results clearly demonstrate the importance of behavior at these scales for observables in large-scale structure. \cite{Debackere:2019cec} has also recently highlighted the importance of this regime for the power spectrum.

To complete our  halo model interpretation of the measured power spectra and equilateral bispectra, we remark that, at a given $k$, each statistic depends on the total mass enclosed within a radius slightly larger than $r \sim k^{-1}$ in different halos, normalized by the mean density.\footnote{In more detail, in Eq.~\eqref{eq:wkm}, the $\sin(kr)/kr$ factor effectively truncates the integral over $4\pi r^2 \rho(r,M)$ at slightly larger than $r\sim k^{-1}$. The $M^{-1}$ in the definition of $W(k,M)$ cancels with the $M$ factors in the integrals in Eqs.~\eqref{eq:p1h}-\eqref{eq:b1h}, leaving a mass-function--weighted integral over some power of $MW(k,M)/\bar{\rho}$.} We plot the ratio of this normalized enclosed mass for the hydro and DMO halos in the bottom row of panels in Fig.~\ref{fig:profiles}. For the lower two mass bins  (which dominate the one-halo term at $3\invMpc \lesssim k \lesssim 10\invMpc$), this ratio for dark matter reaches a maximum at $r\gtrsim 0.5h^{-1}{\rm Mpc}$, roughly translating (within a factor of two in $k$) to the location of the peaks in the power spectrum and bispectrum ratios at $k\sim 3\invMpc$. Similar connections can also be drawn for the baryon and total matter measurements.

Furthermore, we find that the enclosed mass ratio for dark matter is more enhanced higher halo masses, while for baryons it is more suppressed for lower halo masses. Since the bispectrum is more sensitive to higher-mass halos, we would expect that the dark matter bispectrum is more enhanced than the power spectrum, while the baryon power spectrum is more suppressed than the bispectrum, and Fig.~\ref{fig:1halo-comparison} confirms this expectation.

For these comparisons, we defined the halo mass bins separately in the hydro and DMO simulations, rather than attempting to cross-match individual halos between simulations. However, other studies have found that, in the mass range of interest here, halos in hydro simulations can have masses up to $\sim 25\%$ lower than their counterparts in DMO simulations for stronger AGN feedback models (see e.g. Figure 2 of \citealt{Mummery:2017lcn} and Figure 1 of \citealt{Schaller:2014uwa}), corresponding to $\Delta\log(M)\approx -0.1$. We have checked how this might influence our results by re-computing the curves in Fig.~\ref{fig:profiles} assuming that halo masses in the hydro run are systematically low by $\Delta\log(M)\approx -0.1$. While this induces some minor quantitative changes in the shapes of some curves, the noteworthy qualitative features, such as the double-peaked structure of the dark matter curve in the middle-right panel, remain intact. Thus, our conclusions in this section are robust with respect to our method of comparing halos between simulations.

\subsubsection{Non-equilateral bispectrum configurations}

\begin{figure*}
\includegraphics[width=0.985\textwidth]{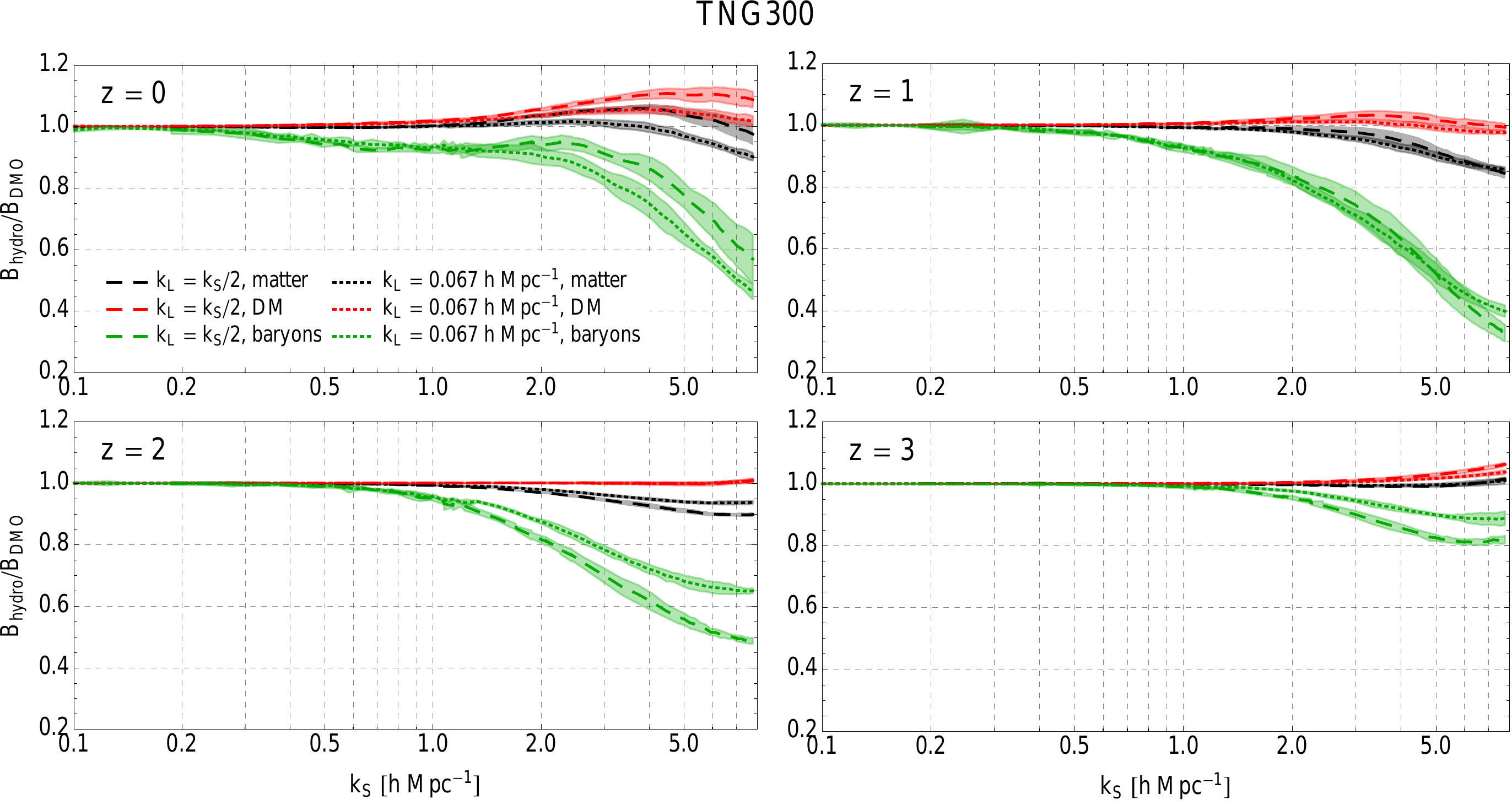}
\caption{\label{fig:tng300-bsisoc}
Same as Fig.~\ref{fig:tng300-psbs}, but for bispectra evaluated on isosceles triangles with two sides with magnitude $\ks$ and the third with magnitude $\kl=\ks/2$ (dashed lines), or isosceles triangles where $\kl$ is set as low as possible such that we can evaluate sample-variance uncertainties with the method of Sec.~\ref{sec:errorbars}. The former bispectra ratios generally track those for equilateral triangles, while the latter (``squeezed") configurations track the corresponding power spectrum ratios (compare with Fig.~\ref{fig:tng300-psbs}).
}    
\end{figure*}

\begin{figure*}
\includegraphics[width=0.985\textwidth]{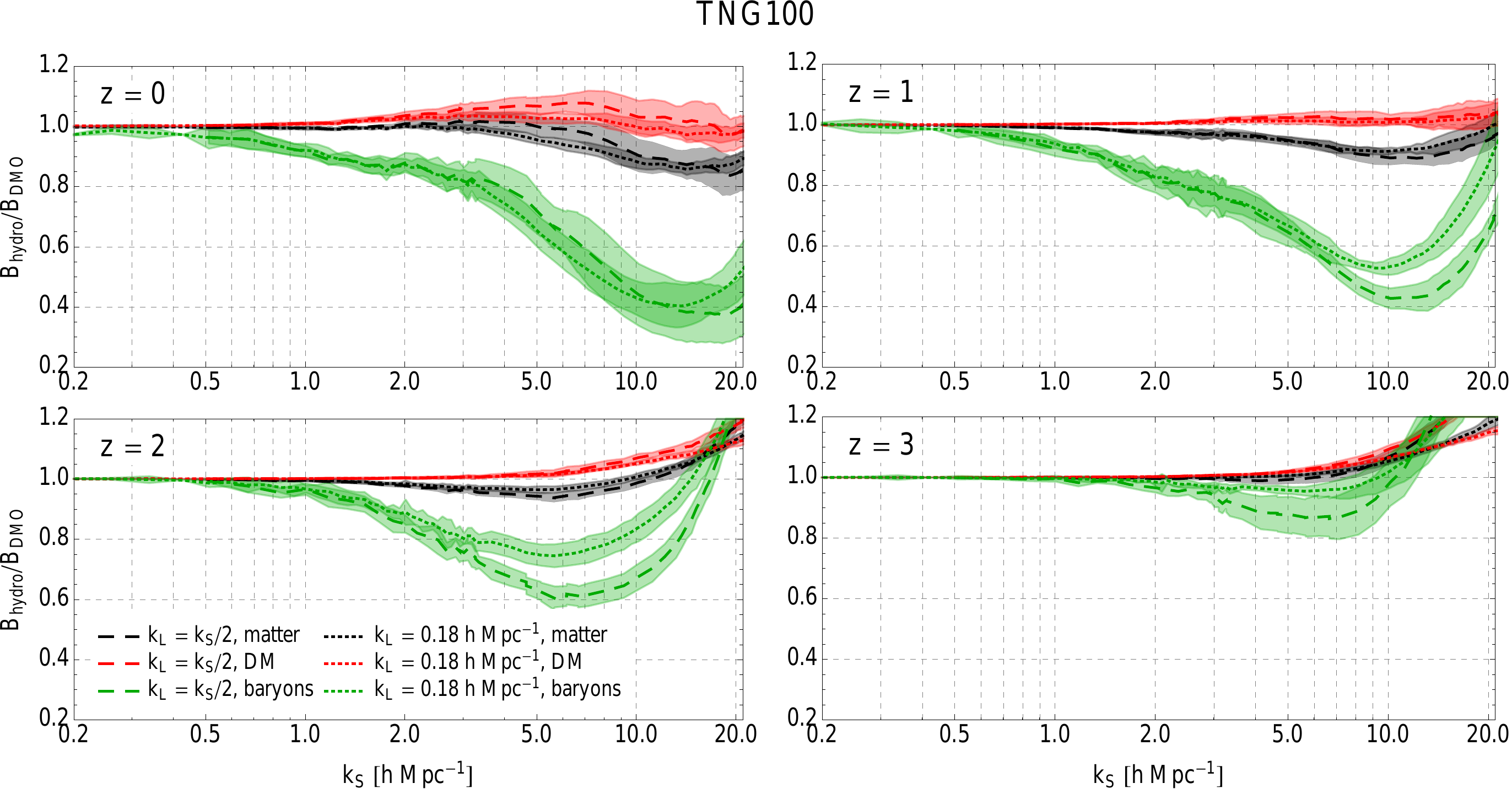}
\caption{\label{fig:tng100-bsisoc}
Same as Fig.~\ref{fig:tng300-bsisoc}, but for measurements from TNG100-1.
}    
\end{figure*}

In Figs.~\ref{fig:tng300-bsisoc} and~\ref{fig:tng100-bsisoc}, we show bispectrum ratios for two isosceles configurations. In the dashed lines, we show the case where $\kl=\ks/2$ (where two triangle sides have length $\ks$ and the third has length $\kl$). In the dotted lines, we fix $\kl$ to the lowest value for which we could estimate sample-variance errorbars from sub-boxes (specifically, $\kl=0.067\invMpc$ and $0.18\invMpc$ for TNG300 and TNG100, respectively); these triangles are mostly ``squeezed," with $\kl\ll \ks$. For $\ks\lesssim 10\invMpc$, the $\kl=\ks/2$ ratios generally track those for the equilateral bispectra, while the squeezed configurations track the associated power spectrum ratios.

These behaviors are not surprising, when we consider that baryonic effects are rather slowly-varying functions of $k$ in this wavenumber range, so configurations with $\kl=\ks/2$ will not experience significantly different effects than equilateral configurations. Furthermore, in a variety of contexts, squeezed bispectra are known to be related to the power spectrum of short modes; in Sec.~\ref{sec:responses}, we compare our measurements to specific predictions for this relationship in large-scale structure.

\subsection{Illustris}
\label{sec:ill}

\begin{figure*}
\includegraphics[width=0.985\textwidth]{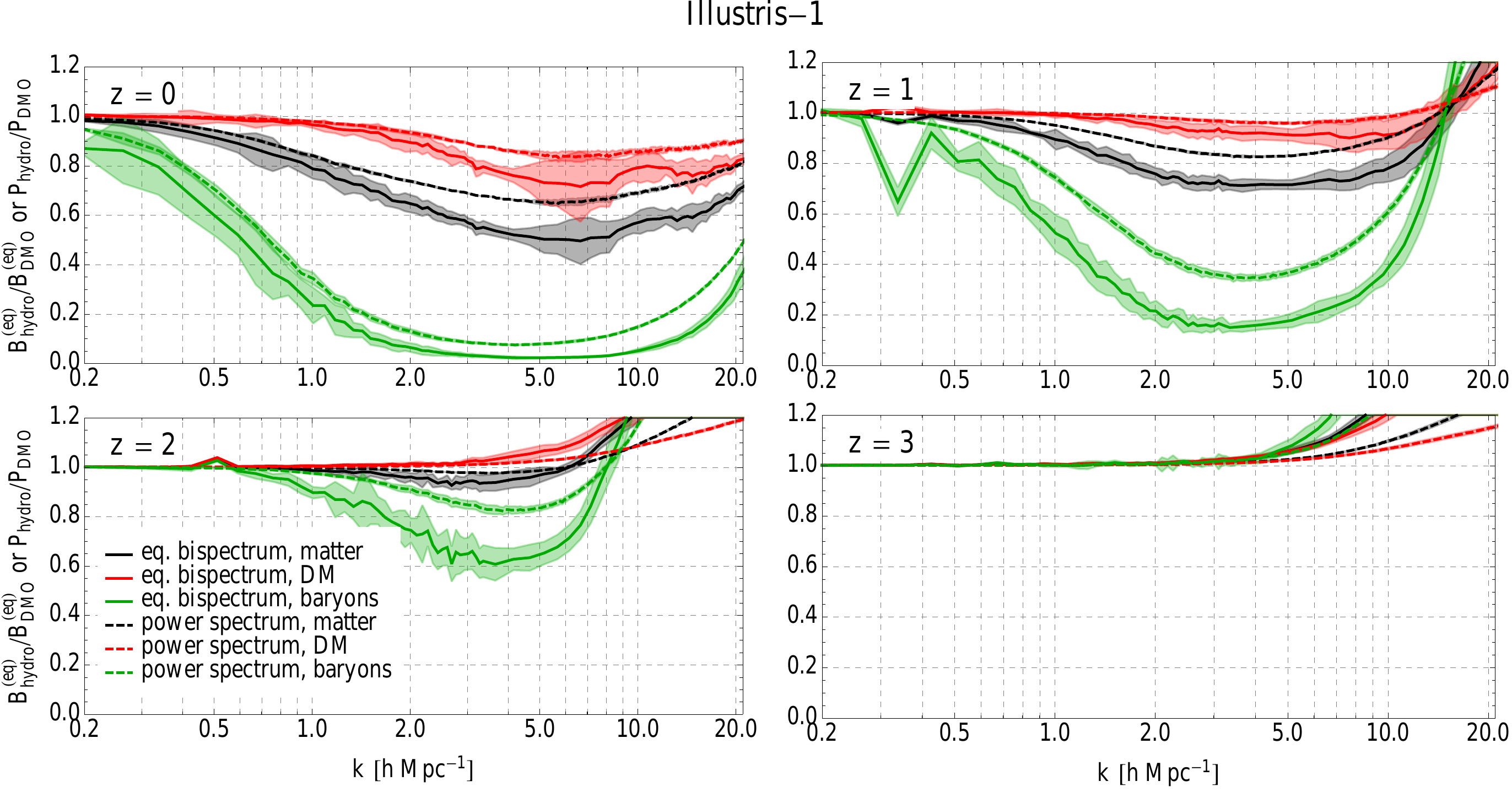}
\caption{
\label{fig:ill-psbs}
Same as Fig. \ref{fig:tng300-psbs} but for the Illustris simulation. All ratios are suppressed much more than in IllustrisTNG,  a manifestation of the extremely efficient AGN feedback implementation in Illustris.}    
\end{figure*}

We also carry out our measurements on the Illustris simulation \citep{Vogelsberger_14a, Vogelsberger:2013eka, Genel_14, Sijacki_15}, in particular on the hydrodynamic Illustris-1 and DMO Illustris-1-Dark runs. The side length of the simulation boxes is $75h^{-1}\,{\rm Mpc}$ and the values of the cosmological parameters are in agreement with the constraints from WMAP9 \citep{Hinshaw:2012aka}. The simulations follow the evolution of $1820^3$ dark matter particles plus $1820^3$ gas elements (only in the hydrodynamic simulation) down to $z=0$. The AREPO code was used to solve the hydrodynamic equations using the Voronoi moving-mesh method~\citep{Springel:2009aa}.

The physics implemented in the hydrodynamic simulation is the following: star formation is simulated using the \cite{Volker_2003} model, supernova feedback is modelled through stellar winds implemented as kinetic outflows, and AGN feedback includes both quasar-mode and radio-mode. The free parameters of the subgrid models have been tuned to reproduce several observations, including the cosmic star formation rate density, stellar mass function, and galaxy luminosity functions in various bands. More details on the subgrid physics implementation in the Illustris simulation can be found in \cite{Vogelsberger:2013eka} and \mbox{\cite{Torrey_2014}}.

In Fig. \ref{fig:ill-psbs}, we show the ratios of the power spectrum and equilateral bispectrum between the hydrodynamic and DMO runs. For baryons, we find similar features to those we observed in the IllustrisTNG simulations: at $z<2$, we find the same spoon-shaped feature as in IllustrisTNG for both the power spectrum and bispectrum. That feature is however shifted to larger scales, because the AGN feedback model implemented in Illustris is more effective than the one in IllustrisTNG and EAGLE (see below). With AGN feedback being able to move large amounts of gas to rather large distances, clustering deviations with respect to the DMO simulation take place on larger scales. Of course, at higher redshifts, this feedback is less effective, and the amplitude of the suppression shrinks. At $z=3$, we observe higher clustering in baryons than in the DMO run, that we do not observe in the IllustrisTNG simulations. This could either signal that some process (such as gas cooling) that condenses baryons at high redshift is more efficient in Illustris, or indicate that some other process such as galactic winds (which are stronger in IllustrisTNG -- see \citealt{Pillepich:2017jle}) expels more baryons at high redshift in IllustrisTNG than Illustris.

For dark matter in the hydro run, we observe a deficit in power (in both the power spectrum and bispectrum) at $z\leq1$, while at higher redshifts we find that dark matter is more clustered  on small scales than matter in the DMO run. We note that, differently than in IllustrisTNG, we do not find an excess of power at low redshift. This is caused by the very effective model of AGN feedback used in Illustris: the backreaction of gas on dark matter is so large that dark matter in the hydro run is also less clustered than total matter in the DMO run. At redshifts $z\geq2$, results from Illustris are similar to those from IllustrisTNG, in agreement with our earlier statement that differences at these redshifts are mostly not driven by AGN feedback.

For total matter, we find similar features as for baryons: a spoon shape at $z\leq2$ and higher clustering than the DMO run at $z=3$. The amplitude of the suppression of the matter clustering (both power spectrum and bispectrum) is lower than for baryons, since dark matter is more strongly clustered than baryons. These features can be understood taking into account the above effects on both gas and dark matter.

\subsection{BAHAMAS and EAGLE}
\label{sec:baheagle}

The BAHAMAS simulations \citep{McCarthy2017} are a set of $400h^{-1}\,{\rm Mpc}$ cosmological hydrodynamic simulations run using a modified version of the Lagrangian TreePM-SPH code Gadget-3 \citep{Springel2005} with $2 \times 1024^3$ particles. The subgrid physics models are those used in the OWLS and cosmo-OWLS papers \citep{Schaye2010,LeBrun2014}. The BAHAMAS simulations calibrate these subgrid models by tuning the subgrid parameters using the present-day galaxy stellar mass function and the hot gas mass fractions of groups and clusters. The idea is that by calibrating these observables to measurements of our Universe, the resulting effects of stellar and AGN feedback on other observables should be more representative.

The calibration approach helps to constrain the relative strengths of supernova and stellar feedback, which are implemented primarily as a kinetic kick to the medium around the supernova, and AGN feedback, which is implemented as a quasar mode (i.e.\ thermal injection once the black hole has stored enough energy to heat the surrounding gas by a temperature $\Delta T_{\rm heat}$). As is explored in \citet{McCarthy2017}, these two feedback modes are coupled as supernova feedback suppresses the efficiency of AGN feedback. Through examining the galaxy stellar mass function, the two effects can be partially disentangled, as they impact the low- and high-mass regimes differently. 

The EAGLE simulation \citep{Schaye2015,Crain:2015,McAlpine:2015tma,EAGLE2017} was also run with a modified version of Gadget-3 but with $2 \times 1504^3$ particles in a $67.7h^{-1}\,{\rm Mpc}$ box. The EAGLE simulation also uses the subgrid models used in the OWLS simulations \citep{Schaye2010} but with several changes. The main changes are that feedback from star formation and supernovae, which are implemented together, is thermal (as opposed to the kinetic implementation in BAHAMAS and OWLS); the accretion of gas onto black holes now accounts for the angular momentum of the gas and, due to the simulation's finer resolution, does not need the density-dependent Bondi accretion enhancement factor; the black hole thermal feedback temperature is much higher ($10^{8.5}\,{\rm K}$, compared to $10^{7.8}\,{\rm K}$ in BAHAMAS); and the star formation density threshold is not a constant as in BAHAMAS, but instead depends on metallicity (motivated by the higher cooling rates of metal-rich areas). The EAGLE simulation was run at the Planck 2013 cosmology \citep{Ade:2013zuv} and, like BAHAMAS, used measurements of the galaxy stellar mass function to calibrate the feedback model parameters. 

\begin{figure*}
\includegraphics[width=0.985\textwidth]{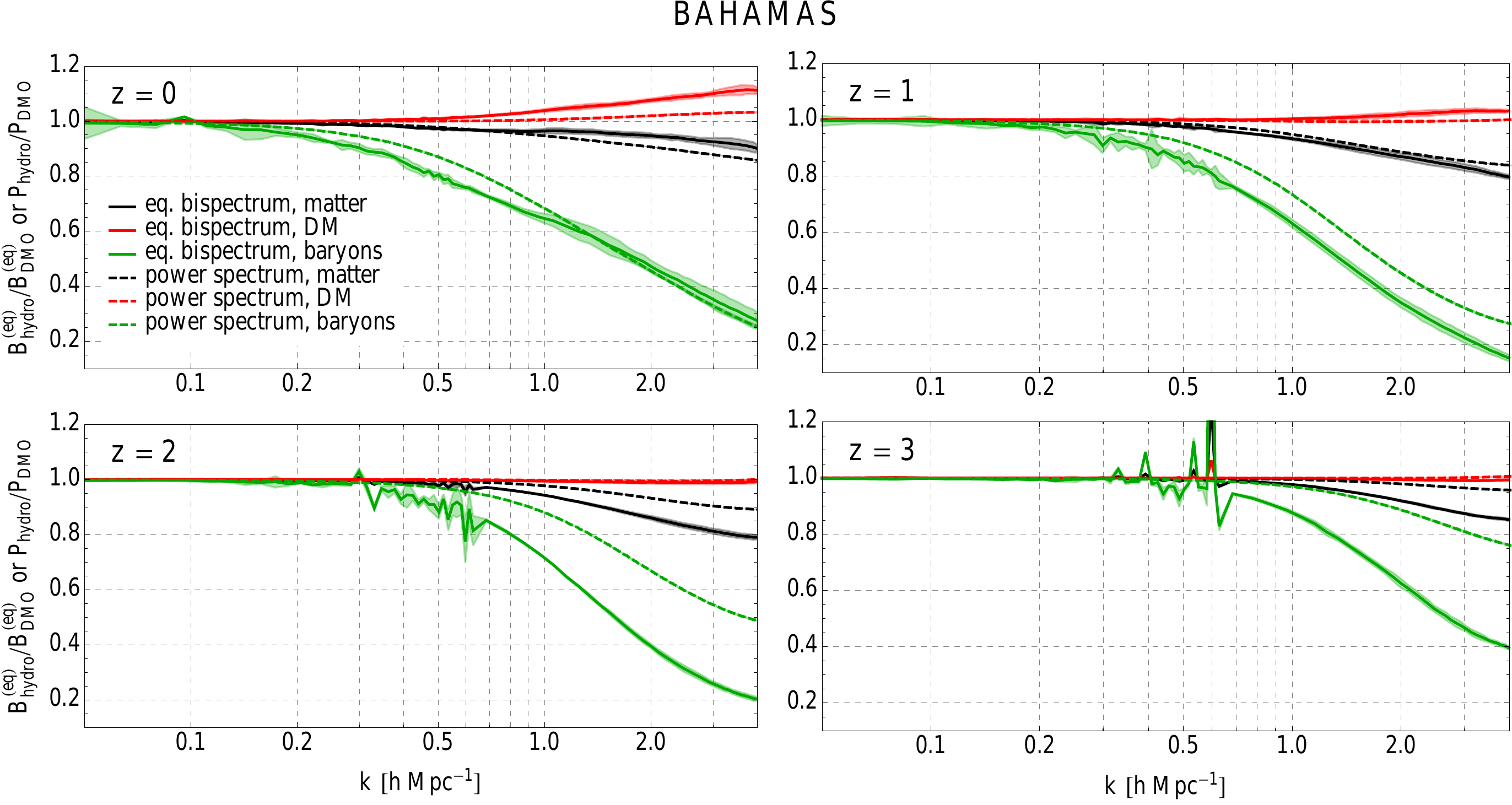}
\caption{
\label{fig:bah-psbs}
Same as Fig.~\ref{fig:tng300-psbs} but for the BAHAMAS simulation with the fiducial AGN feedback prescription. Compared to IllustrisTNG (see Figs.~\ref{fig:tng300-psbs} and~\ref{fig:tng100-psbs}) the spectra are more strongly suppressed at high redshift, likely due to a combination of gas cooling and feedback.
}
\end{figure*}

\begin{figure*}
\includegraphics[width=0.985\textwidth]{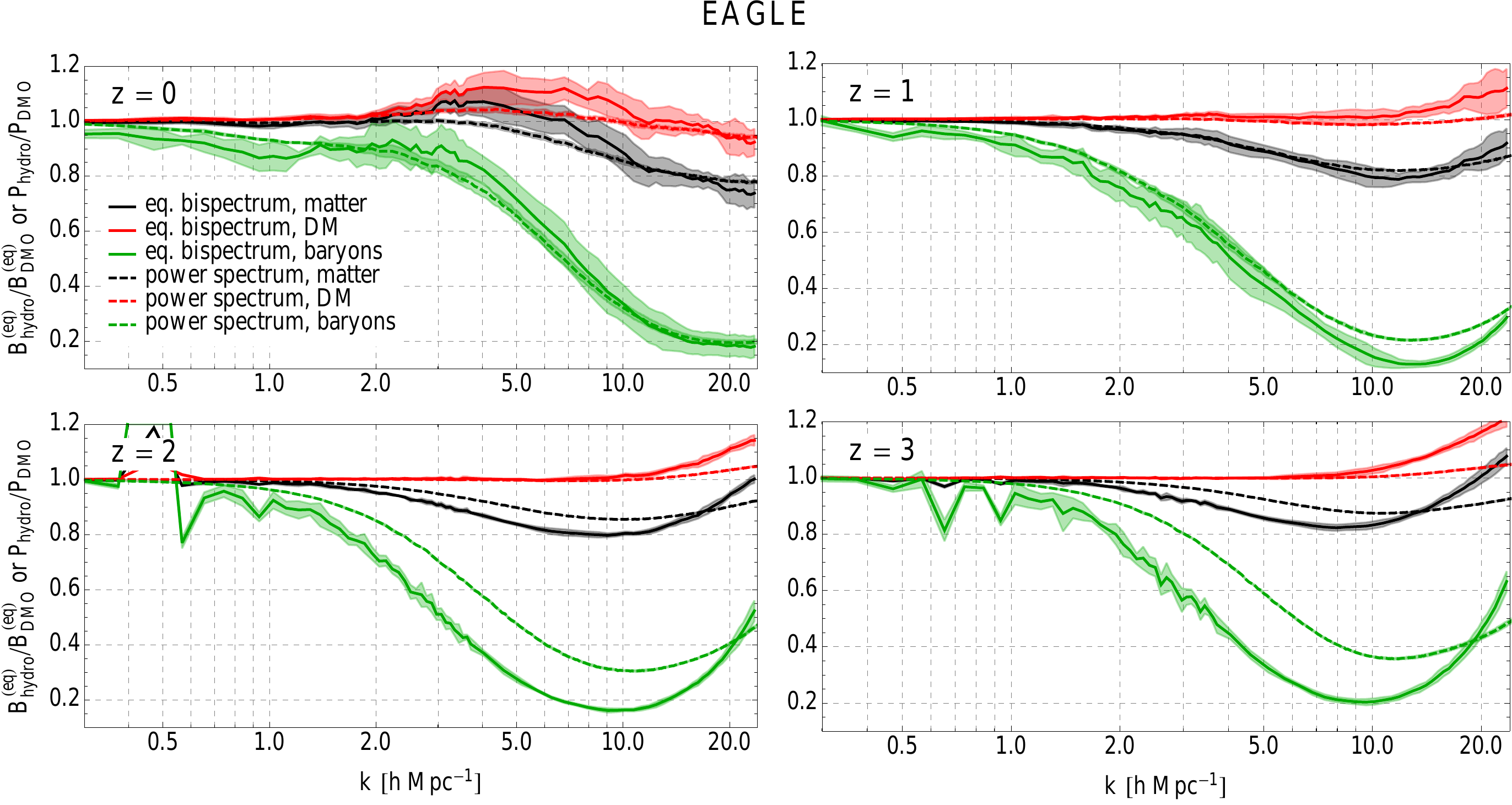}
\caption{
\label{fig:eagle-psbs}
Same as Fig. \ref{fig:tng300-psbs} but for the EAGLE simulation.
}
\end{figure*}

In \citet{Mccarthy2018}, the BAHAMAS suite of simulations was extended to include simulations with AGN injection temperatures that are $0.2\,{\rm dex}$ higher and lower than the fiducial value of $10^{7.8}\,{\rm K}$. In this work, we consider all three feedback models, hereafter referred to as the fiducial, low-AGN ($-0.2$ dex) and high-AGN ($+0.2$ dex) models. This range of AGN temperatures is a measure of the degree of uncertainty in the modelling of the baryonic processes. We use the BAHAMAS simulations that were run with the WMAP9 cosmology \citep{Hinshaw:2012aka}. We note that, as was shown in \cite{2015MNRAS.454.1958M,Mummery:2017lcn,vanDaalen:2019pst}, the fractional effect of feedback (with respect to the DMO runs) is only weakly dependent on cosmology, and thus the difference in the WMAP9 and Planck 2013 cosmologies will not significantly affect our conclusions.

In Figure~\ref{fig:bah-psbs}, we plot ratios of the power spectrum and equilateral  bispectrum for the fiducial BAHAMAS model, and in Figure~\ref{fig:eagle-psbs} we plot the same for the EAGLE simulations. At $z=3$, we find more suppression of the small scale power spectrum and bispectrum for all species (baryons, dark matter and total matter), compared to either Illustris or IllustrisTNG. This is also seen in the Horizon-AGN simulations \citep{Chisari:2018prw}. This is explained by a combination of varying efficiencies of gas cooling and AGN feedback at high redshift in these simulations. Specifically, in Illustris and IllustrisTNG, the effects of adiabatic cooling, and consequent boost on the collapse of matter, are more visible at high redshift because of the weaker strength of AGN feedback at those redshifts in these simulations. In BAHAMAS and EAGLE, on the other hand, AGN feedback at high redshift is strong enough to mostly outweigh the effects of adiabatic cooling on the scales shown (for EAGLE, at $z = 3, k \gtrsim 10\invMpc$ one can note an upturn in the curves, indicative of the growing importance of gas cooling on smaller scales). The difference in early feedback strengths can be related to the specific subgrid recipes: in IllustrisTNG, the continuous nature of thermal AGN feedback leads the resulting thermal energy to rapidly radiate away for numerical reasons, whereas the stochastic implementation in BAHAMAS and EAGLE prevents such an escape and leads to a stronger effect~\citep{Davies:2019arxiv,Weinberger:2017bbe}.

For both BAHAMAS and EAGLE, the suppression in both the equilateral bispectrum and power spectrum between $z=3$ and $z=1$ does not evolve as much as in Illustris and IllustrisTNG. This holds generically for baryons, dark matter and total matter, and it reflects the growing strength of AGN feedback in Illustris and IllustrisTNG during these epochs. For EAGLE and BAHAMAS, on the other hand, the stronger early time suppression leads to there being less matter density on small scales to fuel AGN feedback at later times, effectively decreasing its efficiency and impact on the clustering of matter.

Between $z=1$ and $z=0$, the results for both BAHAMAS and EAGLE display an enhancement in the amplitude of the power spectrum and equilateral bispectrum of dark matter. This is similar to what happens in other simulations (with the exception of Illustris) and that we attempted to describe already above in Sec.~\ref{sec:illustrisTNG}: closer to the present epoch, the reduced efficiency of AGN feedback allows previously expelled matter to re-collapse into the halos. The imprint of this re-collapse can be seen in the dark matter component as a backreaction.  This idea is also supported by the results in \citet{Chisari:2019tus}, who find that this dark matter backreaction excess at $k \sim 2\invMpc$ does not occur in their AGN-free simulation, suggesting that it requires material to be heated or ejected by the AGN for this feature to form. 

\begin{figure*}
\includegraphics[width=0.985\textwidth]{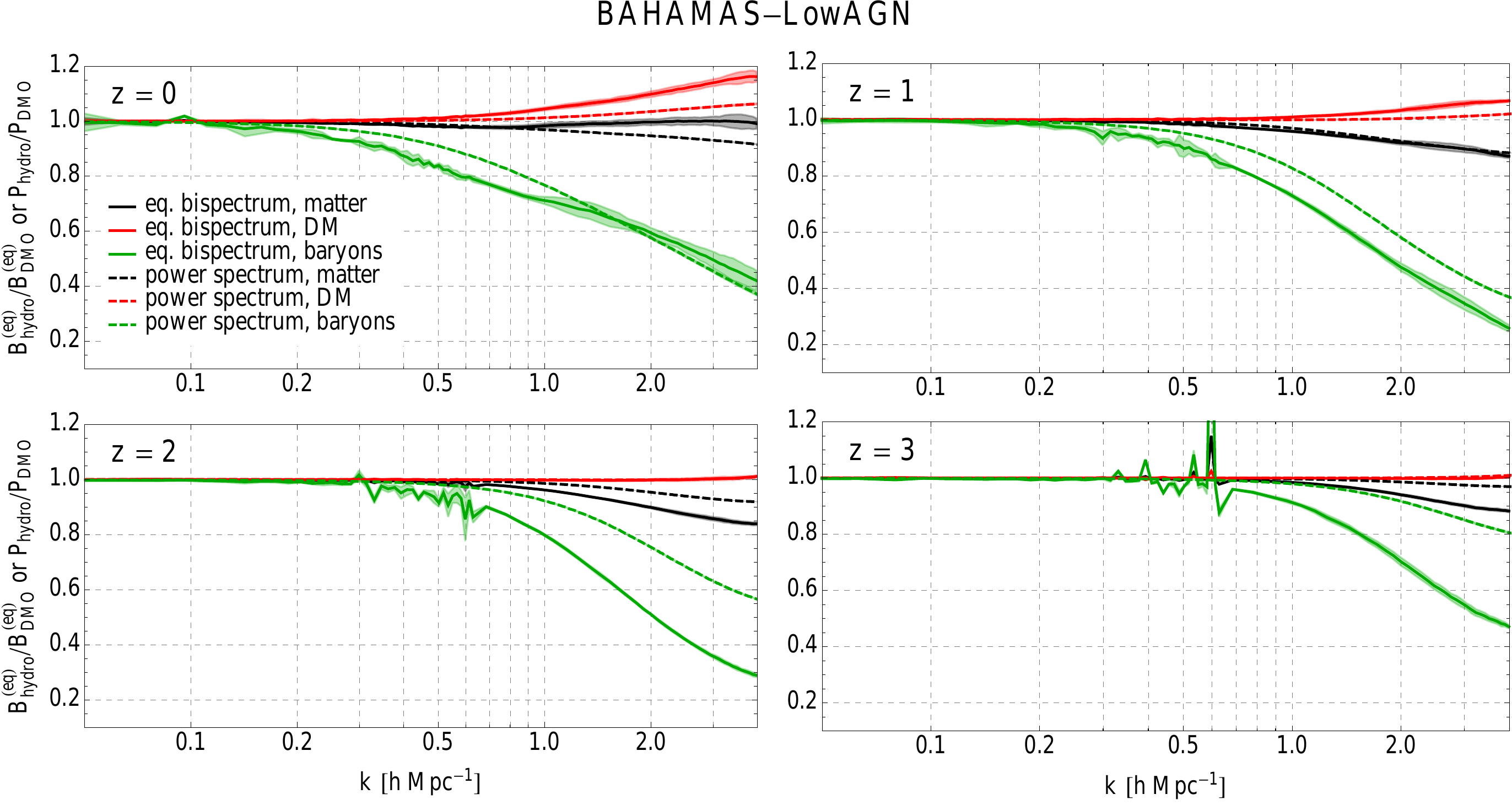}
\caption{
\label{fig:bah-lowAGN-psbs}
Same as Fig. \ref{fig:tng300-psbs} but for the BAHAMAS simulation with the ``low-AGN" feedback prescription. As expected, weaker feedback results in weaker power suppression compared to the DMO run.
}    
\end{figure*}

\begin{figure*}
\includegraphics[width=0.985\textwidth]{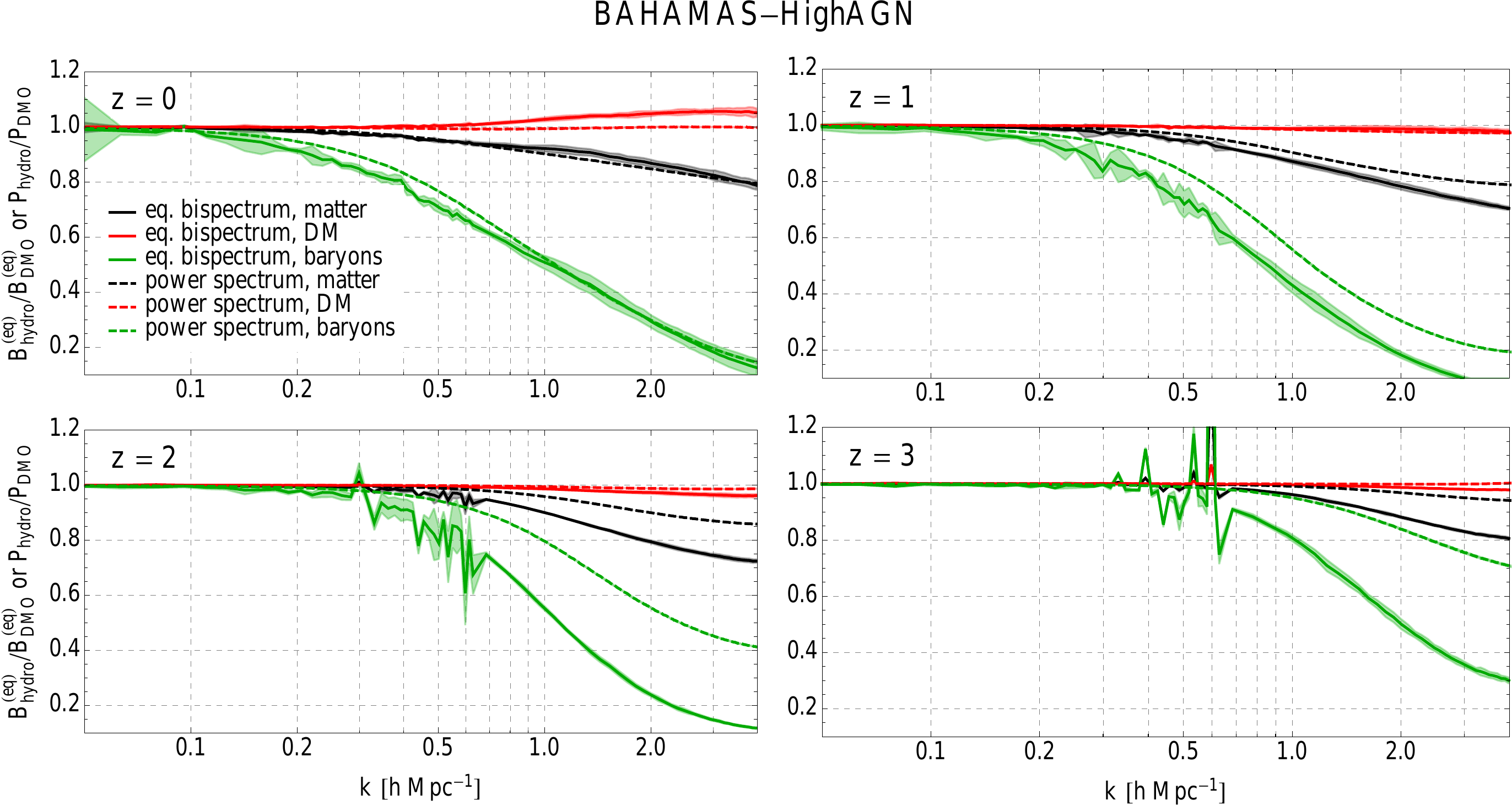}
\caption{
\label{fig:bah-hiAGN-psbs}
Same as Fig. \ref{fig:tng300-psbs} but for the BAHAMAS simulation with the ``high-AGN" feedback prescription.
}    
\end{figure*}

The corresponding results for the low-- and high--AGN-temperature variants of BAHAMAS are shown in Figs.~\ref{fig:bah-lowAGN-psbs} and~\ref{fig:bah-hiAGN-psbs}. Comparing with the fiducial model, we find that the general effect of increasing the feedback temperature is to increase the suppression for baryons and total matter and to reduce the backreaction on the dark matter. With higher feedback temperatures, the gas will expand to larger radii; however, we note that the higher temperature also means longer duty cycles and so more sporadic feedback. This is a feature of how the feedback is implemented, with the AGN storing a fraction of its accreted energy to release as feedback. For higher feedback temperatures, more energy has to be accumulated. However, there is also a dependence on mass resolution: despite the feedback temperature being a factor of 10 higher in EAGLE than in BAHAMAS, the latter shows more $z=0$ power suppression at $k<3\invMpc$. The lower mass resolution of BAHAMAS implies that, per heating event, more thermal energy is deposited in BAHAMAS than in EAGLE, leading to stronger expulsion of the nearby gas overall.

Common to all simulations analyzed in this paper, the effects on the bispectrum are generically stronger than in the power spectrum. Specifically, the fractional suppression for baryons and total matter at higher redshifts is more significant in the bispectrum, and the positive backreaction on the dark matter component is also more noticeable in its bispectrum than in the power spectrum. This is explored in more detail below in Sec.~\ref{sec:modellingApproaches}, and again indicates that the bispectrum can provide highly complementary information to the power spectrum.

Note that \cite{vanDaalen:2019pst} has found that, for $k\leq 1\invMpc$, suppression of the matter power spectrum at a given $k$ can be directly linked to the mean baryon fraction in halos with mass $\sim 10^{14}M_\odot$. It would be interesting to see if such a link can be found for the total matter bispectrum, and also for the separate dark matter and baryon bispectra. We leave this to future work.

\subsection{Summary}
\label{sec:summary}

\begin{figure}
\includegraphics[width=\columnwidth]{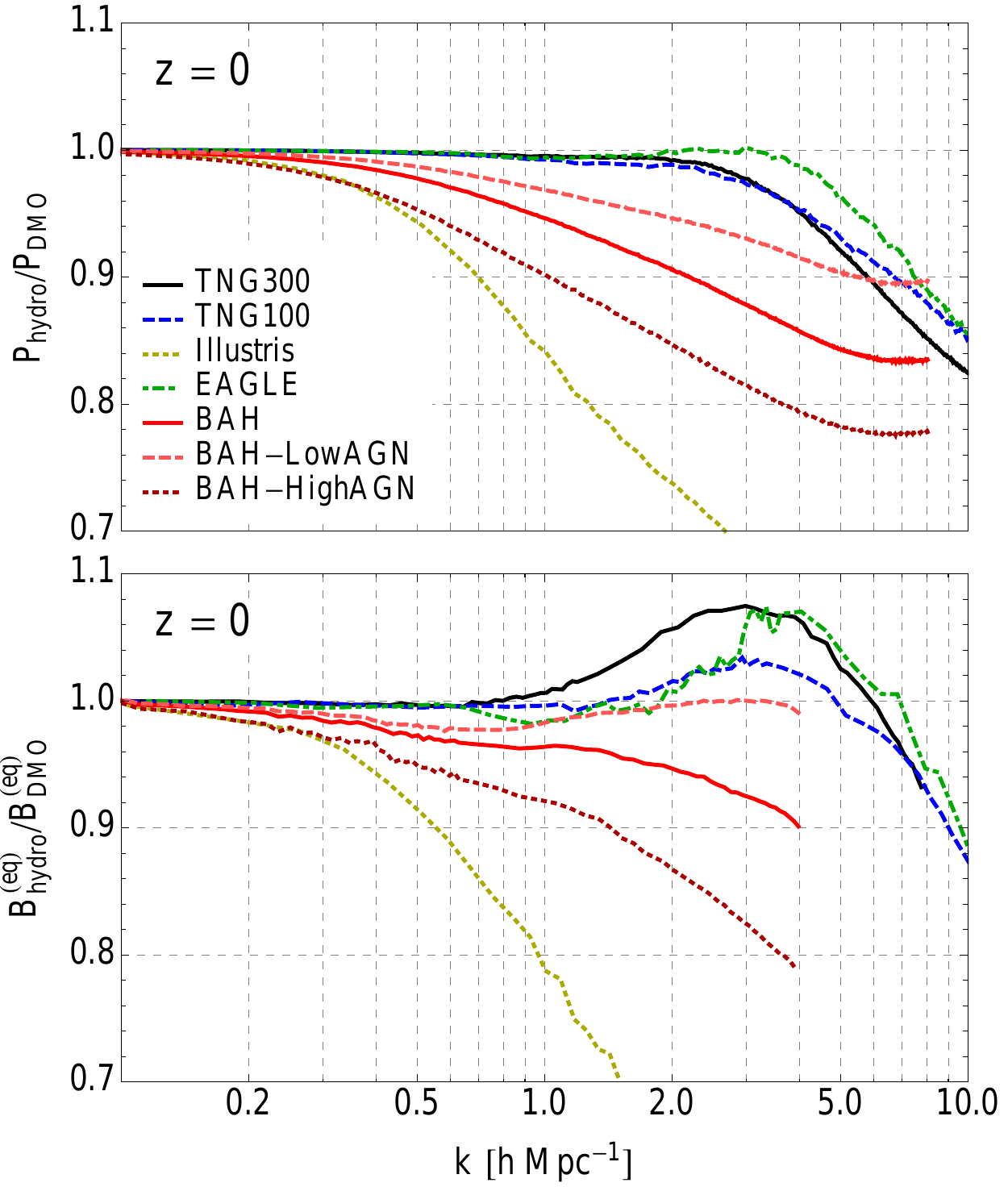}
\caption{\label{fig:summary}
Summary of our measurements of matter power spectra (upper panel) and equilateral matter bispectra (lower panel) at $z=0$, each normalized to the corresponding DMO measurement. To avoid clutter, we do not show sample-variance uncertainties on each ratio. The varying levels of suppression are mostly due to different implementations of AGN feedback, while enhancements of the bispectrum are due to backreaction of baryons on dark matter within massive halos.
}    
\end{figure}

In Fig.~\ref{fig:summary}, we show all measurements of matter power spectra (upper panel) and equilateral matter bispectra (lower panel) we have made at $z=0$, each normalized to the corresponding DMO measurement. Varying levels of power suppression can be seen, dominated by differences between implementations of AGN feedback, while simulations with somewhat weaker feedback exhibit enhancements of their bispectra peaking around $k\sim 3\invMpc$; we have argued that these enhancements are due to baryons backreacting on the underlying distribution of dark matter in massive halos.

%--------------------------------------------------------------------------------------
% MATTER BISPECTRUM AT LARGE SCALES
%--------------------------------------------------------------------------------------
\section{Matter bispectrum at large scales}
\label{sec:largescales}

\begin{figure*}
\includegraphics[width=0.99\textwidth]{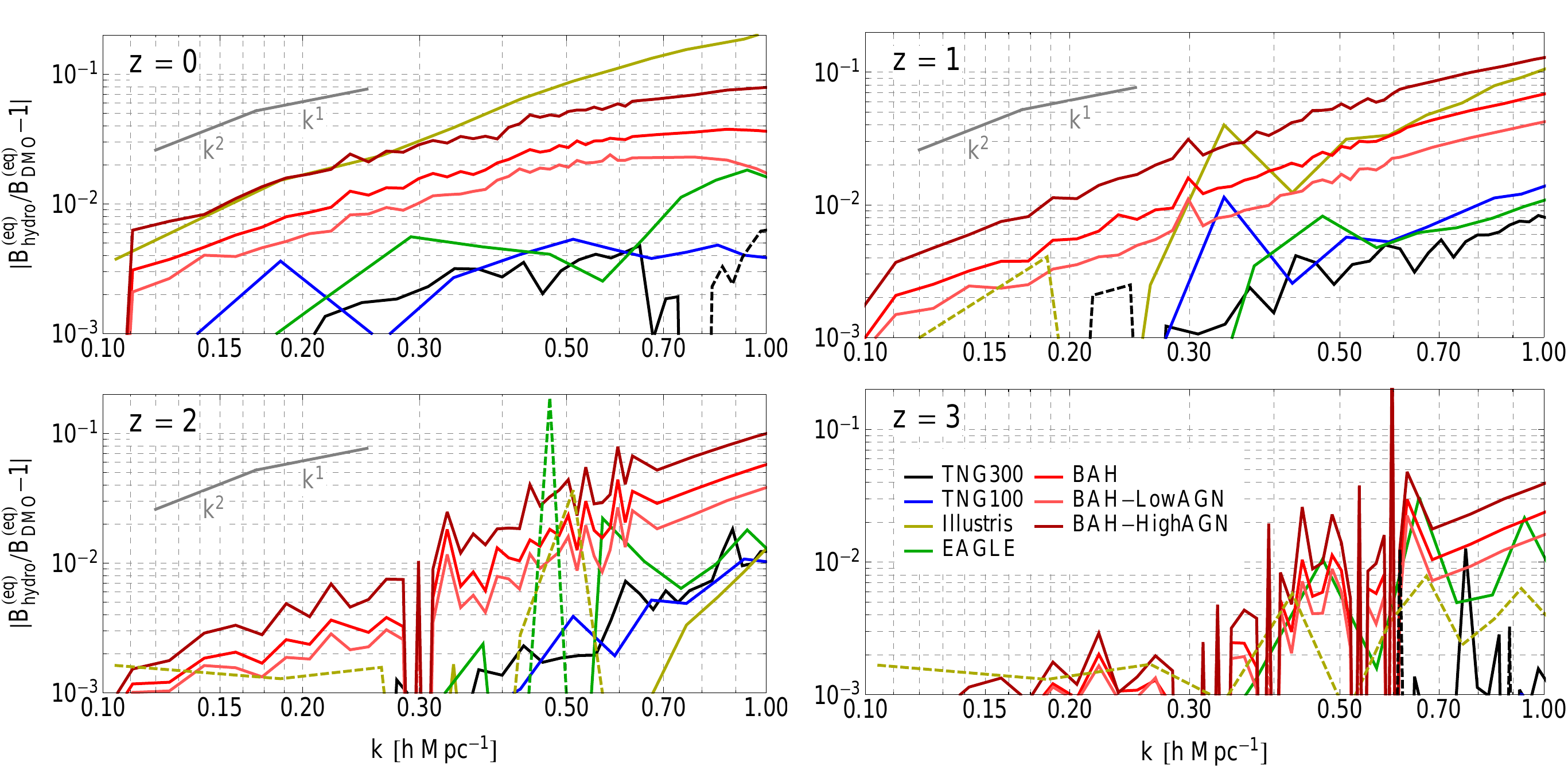}
\caption{
\label{fig:bs-lowk}
Fractional difference between equilateral total-matter bispectra in hydro and DMO runs of each simulation suite we consider in this work. (For clarity, we do not plot the associated sample-variance uncertainties.) For the three BAHAMAS simulations, the difference can approach the level at which the bispectrum changes in the presence of massive neutrinos ($\sim$5\% between $\sum m_\nu = 0$ and $0.06\,{\rm eV}$), while the difference is much smaller for IllustrisTNG and EAGLE. In grey, we show power laws scaling like $k^2$ and $k^1$. The fractional bispectrum difference scales roughly like $k^2$ over a range of scales that increases with redshift, roughly matching the expectation from perturbation theory~\protect\citep{Lewandowski:2014rca}.
}    
\end{figure*}

It is useful to examine baryonic effects on the bispectrum on relatively large scales ($k\lesssim 1\invMpc$), since this is the region where theoretical modelling is most under control, in principle enabling tighter constraints on expansion history~\citep{Gil-Marin:2014sta,Gil-Marin:2016wya}, neutrino mass~\citep{Coulton:2018ebd,Chudaykin:2019ock,Hahn:2019zob}, and primordial non-Gaussianity~\citep{Karagiannis:2018jdt} than obtainable from the power spectrum alone. In Fig.~\ref{fig:bs-lowk}, we show the fractional difference between equilateral total-matter bispectra in hydro and DMO runs of each simulation suite we consider. For IllustrisTNG and EAGLE, the difference is mostly below 1\%, while for BAHAMAS it exceeds 1\% at $z=0$ for $k\gtrsim 0.3$, $0.22$, and $0.15\invMpc$ for the LowAGN, fiducial, and HighAGN versions respectively.

For BAHAMAS, these differences can approach the amount by which the bispectrum changes in the presence of massive neutrinos, roughly 5\% between $\sum m_\nu = 0$ and $0.06\,{\rm eV}$~\citep{Ruggeri:2017dda,Hahn:2019zob}. However, given the different redshift- and scale-dependence of baryonic and neutrino-mass effects (as investigated for the matter power spectrum in~\citealt{Mummery:2017lcn}), it is likely that the two can be disentangled effectively, although this remains to be quantified for the bispectrum. On the other hand, for the implementations of the IllustrisTNG and Eagle simulations, baryonic effects are less of an issue for constraining other physics with the bispectrum.

All fractional differences (except for those curves with too much scatter to draw conclusions about their shapes) appear to be well-described by slowly-varying power laws in wavenumber. For visual comparison, in Fig.~\ref{fig:bs-lowk} we also display lines corresponding to $k^2$ and $k$ power laws. The $k^2$ case is particularly interesting, as such behavior appears within the effective field theory approach to perturbation theory as a consequence of short-distance, nonlinear processes backreacting on large-scale clustering patterns~\citep{Carrasco:2012cv,Carrasco:2013mua}. In particular,~\cite{Lewandowski:2014rca} derived the leading baryonic effects on the large-scale matter power spectrum in this context, and their result implies that the fractional power spectrum difference between including or ignoring these effects scales like $k^2$; further, they found good agreement between this prediction and a variety of power spectra from the OWLS simulations~\citep{2011MNRAS.415.3649V}. 

In many cases, our measured equilateral bispectra show this behavior, with the fractional difference between hydro and DMO measurements scaling roughly like $k^2$. The tilt generally becomes shallower at higher $k$, while higher redshifts have a greater wavenumber range where $k^2$ seems more accurate. It would be interesting for future studies to perform a more detailed comparison between perturbative approaches such as \cite{Lewandowski:2014rca} and \mbox{\cite{Chen:2019cfu}}, extended to bispectra (e.g.\ \citealt{Angulo:2014tfa,Baldauf:2014qfa}),  and the measured power spectra and bispectra from these simulations.

%--------------------------------------------------------------------------------------
% COMPARISONS TO MODELLING APPROACHES
%--------------------------------------------------------------------------------------
\section{Comparisons to modelling approaches} 
\label{sec:modellingApproaches}

In this section, we compare our bispectrum measurements to two modelling approaches: a commonly-used fitting formula for the matter bispectrum (Sec.~\ref{sec:scgm-fitfunctions}), and predictions for squeezed triangles based on nonlinear response functions (Sec.~\ref{sec:responses}).

\subsection{Comparison to bispectrum fitting formulas}
\label{sec:scgm-fitfunctions}

\cite{Scoccimarro:2000ee} present a fitting formula for the matter bispectrum based on ``hyperextended perturbation theory"~\citep{Scoccimarro:1998uz}, designed to interpolate between tree-level standard perturbation theory (e.g.~\citealt{Bernardeau:2001qr}) at large scales and observed behavior of hierarchical clustering amplitudes (e.g.\ skewness) at small scales. This interpolation is accomplished by a phenomenological modification of the symmetrized mode-coupling kernel $F_2(\vk_1,\vk_2)$ from perturbation theory; the fitting formula then reads
\beq
B(\vk_1,\vk_2,\vk_3) = 2F_2^{\rm (eff)}(\vk_1,\vk_2) P(k_1) P(k_2) + \text{2 perms}\ ,
\label{eq:bsc}
\eeq
where $P(k)$ is the nonlinear matter power spectrum, the precise form of $F_2^{\rm (eff)}$ is given in~\cite{Scoccimarro:2000ee}, and all quantities have an implicit redshift-dependence. This formula has been widely used in the literature: for example,~\cite{MacCrann:2016gac} have used it to estimate the size of reduced-shear corrections to the cosmic shear power spectrum, and~\cite{Namikawa:2018erh} have modified it for scalar-tensor theories of gravity. A more recent update of its parameters and fitting form~\citep{GilMarin:2011ik} has also been used to analyze the galaxy bispectrum from BOSS~\citep{Gil-Marin:2014sta,Gil-Marin:2016wya} and model the bispectrum of CMB lensing and associated systematics~\citep{Namikawa:2016jff,Ferraro:2017fac,Bohm:2018omn}
In particular, the original formula was used in~\cite{Semboloni:2012yh} to derive a halo model to describe baryonic effects on weak-lensing observables, and was also used by~\cite{Fu:2014loa} to argue that baryonic effects imparted no more than a 5\% bias to their lensing-based cosmological constraints.

\begin{figure*}
\includegraphics[width=\textwidth]{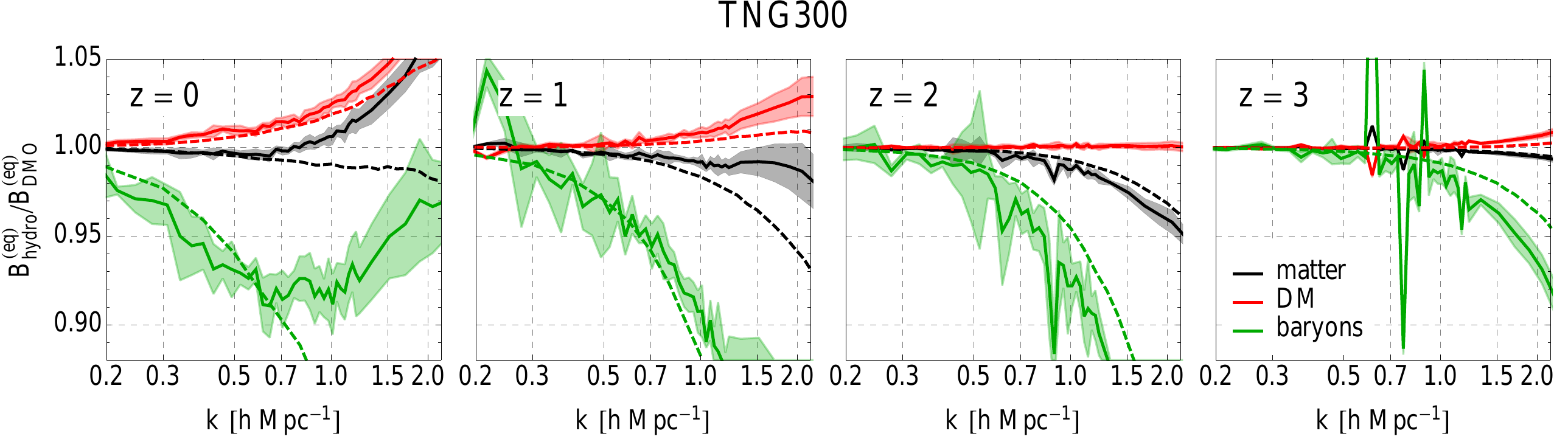}
\caption{\label{fig:tng300-fitfunc}
Comparison of Eq.~\eqref{eq:fitfunc-eq} (dashed lines), derived from the bispectrum fitting function from~\protect\cite{Scoccimarro:2000ee}, 
to equilateral bispectrum measurements from TNG300 (solid lines). The fitting function agrees with the measured bispectrum ratios within our estimated errorbars for $k\lesssim 0.7\invMpc$ at $z\leq 1$, with better performance for dark matter and total matter at higher redshift. 
Up to these rather nonlinear scales, the bispectrum ratios are therefore completely described by the corresponding power spectrum ratios [which are all that enter Eq.~\eqref{eq:fitfunc-eq}].
}    
\end{figure*}

\begin{figure*}
\includegraphics[width=\textwidth]{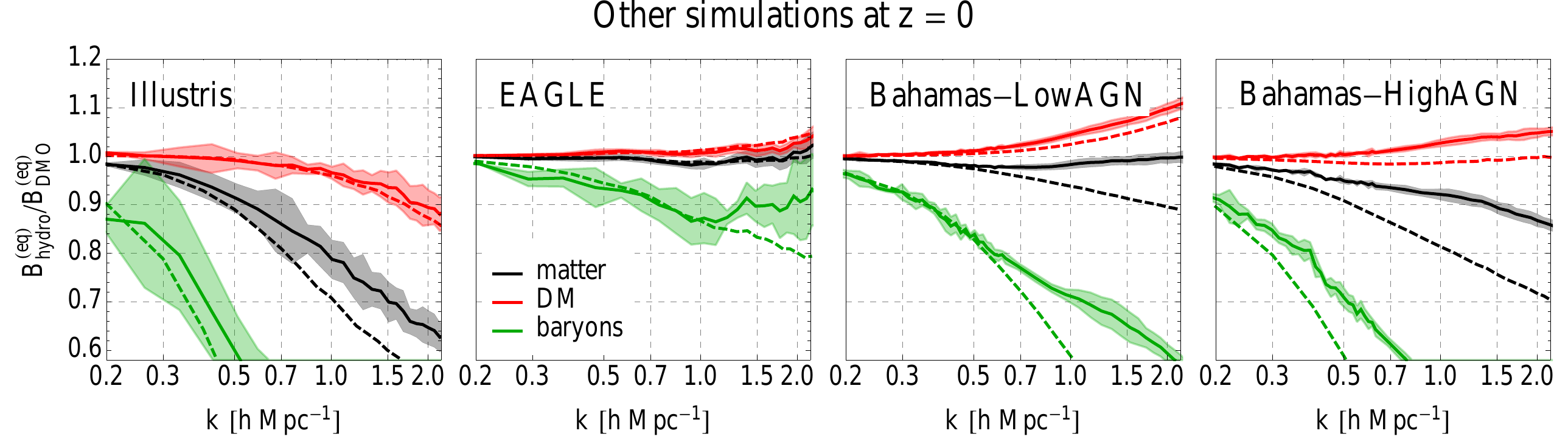}
\caption{\label{fig:othersims-fitfunc-z0}
Same as Fig.~\ref{fig:tng300-fitfunc}, but for other simulations at $z=0$. Eq.~\eqref{eq:fitfunc-eq} generally performs better for simulations for which the power spectrum is more mildly suppressed at the plotted scales, corresponding to weaker overall AGN feedback.
}    
\end{figure*}

With our bispectrum measurements, we are able to re-examine the use of Eq.~\eqref{eq:bsc} to describe the matter bispectrum including baryonic effects. Specifically, we ask whether this formula can describe the {\em relative} effect of baryons on various bispectra, in the form of the ratios from Sec.~\ref{sec:measurements}. We will also push further than this, in two ways: we will examine the validity of the formula beyond the range over which~$F_2^{\rm (eff)}$ was originally fit ($k\leq0.4\invMpc$ in \citealt{GilMarin:2011ik}), and we will ask how it performs when applied to separate bispectra of dark matter and baryons (even though it was originally fit to DMO simulations). We will see that the formula is surprisingly effective in describing the bispectrum ratios in all cases, irrespective of the precise form of~$F_2^{\rm (eff)}$, because the bispectrum ratios are almost completely determined by ratios of the corresponding power spectra

\begin{figure*}
\includegraphics[width=\textwidth]{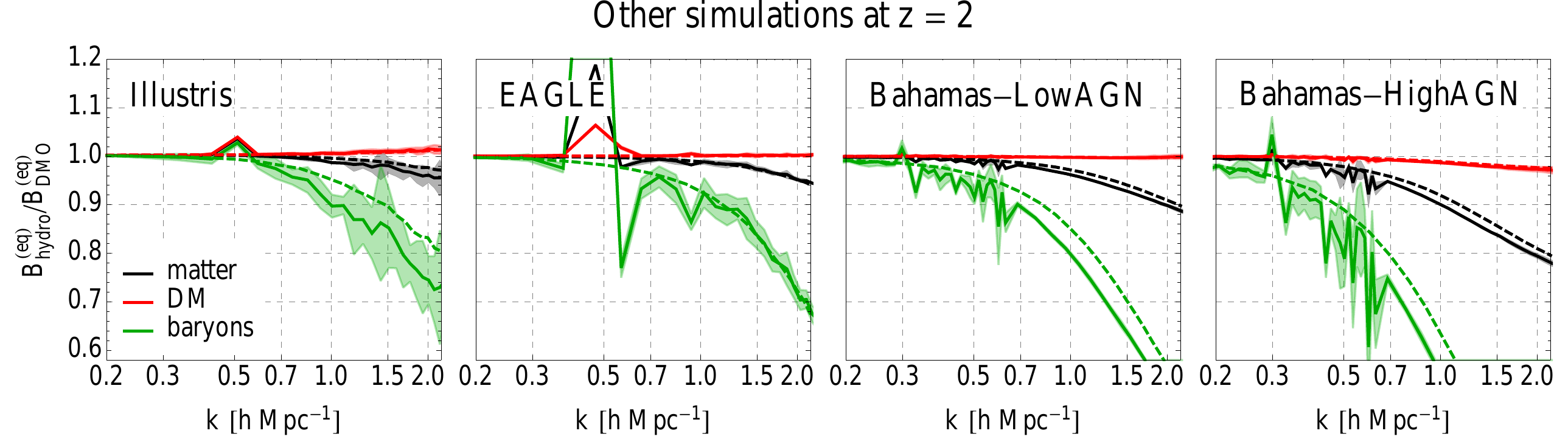}
\caption{\label{fig:othersims-fitfunc-z2}
Same as Fig.~\ref{fig:othersims-fitfunc-z0}, but at $z=2$.
}    
\end{figure*}

\begin{figure*}
\includegraphics[width=\textwidth]{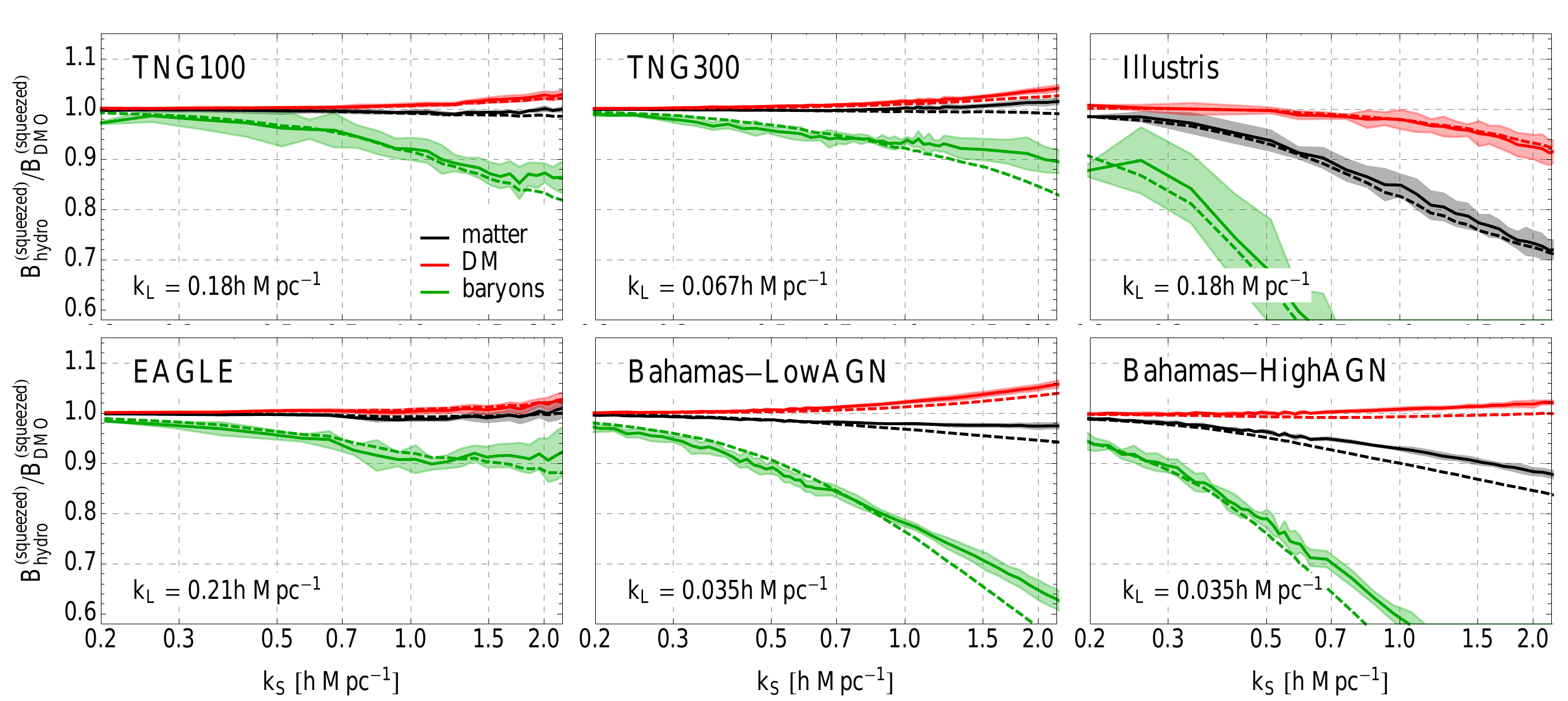}
\caption{\label{fig:allsims-fitfunc-squeezed-z0}
Comparison of Eq.~\eqref{eq:fitfunc-isos} (dashed lines), derived from the bispectrum fitting function from~\protect\cite{Scoccimarro:2000ee}, to squeezed bispectrum measurements from various simulations at $z=0$ (solid lines). The fitting function matches the measurements up to much higher wavenumbers than for equilateral configurations (compare with Figs.~\ref{fig:tng300-fitfunc} and~\ref{fig:othersims-fitfunc-z0}), which can be understood by examining the form of Eq.~\eqref{eq:fitfunc-isos} (see main text for details).
}    
\end{figure*}

We first consider equilateral bispectra. From Eq.~\eqref{eq:bsc}, if we assume that the mode-coupling kernel $F_2^{\rm (eff)}$ is identical for each particle species (DMO, total matter, dark matter, and baryons), the fitting-formula expressions for these ratios have no dependence on $F_2^{\rm (eff)}$, leaving a simple relationship between the equilateral bispectrum and power spectrum ratios:
\beq
\frac{B_X^{\rm (eq)}(k)}{B_{\rm DMO}^{\rm (eq)}(k)} 
	= \lp \frac{P_X(k)}{P_{\rm DMO}(k)} \rp^2\ .
	\label{eq:fitfunc-eq}
\eeq
In Fig.~\ref{fig:tng300-fitfunc}, we compare Eq.~\eqref{eq:fitfunc-eq} with measurements from TNG300 (the comparison with TNG100 is similar). We find that Eq.~\eqref{eq:fitfunc-eq} matches the measured bispectrum ratio for dark matter, baryons, and total matter within our estimated errorbars for $k\lesssim 0.7\invMpc$ at $z\leq 1$, with better performance for dark matter and total matter at higher redshift.\footnote{We note in passing that better matches to the measurements can generally be obtained by fitting for the power in Eq.~\eqref{eq:fitfunc-eq} instead of fixing it to~2. However, such a procedure has no obvious physical justification, so we do not pursue it further.} Thus, up to somewhat nonlinear scales, the influence of baryons on these bispectra is determined almost completely by the relative changes in the power spectra.

We also compare Eq.~\eqref{eq:fitfunc-eq} to equilateral bispectra from other simulations at $z=0$ (Fig.~\ref{fig:othersims-fitfunc-z0}) and $z=2$ (Fig.~\ref{fig:othersims-fitfunc-z2}). The formula generally performs better for simulations with milder power spectrum suppressions at large scales. The match with EAGLE is particularly impressive, with the formula falling within the errorbars on the total matter and dark matter bispectra for $k\lesssim 2\invMpc$ and the baryon bispectrum for $k\lesssim 1.2\invMpc$ at $z=0$. The formula agrees with BAHAMAS-LowAGN at $k\lesssim 0.5\invMpc$ at $z=0$, while agreement with the HighAGN version is much poorer, ceasing (based on where the formula falls outside the errorbars) around $k\sim 0.2\invMpc$. Despite having even stronger feedback, the performance for Illustris is somewhat better than for BAHAMAS-HighAGN.

It is worth noting that at $z=2$, Eq.~\eqref{eq:fitfunc-eq} generally matches the shape of the equilateral bispectrum ratio to within 10\%, so for estimates of baryonic effects at this redshift and higher, it will likely provide a useful guideline (depending on the application).

For non-equilateral configurations, the $F_2^{\rm (eff)}$ kernel does not immediately cancel out in the hydro/DMO ratio as it does for equilateral triangles, so we must use the full expression from Eq.~\eqref{eq:bsc} in the numerator and denominator of the ratio. This expression performs similarly to the equilateral case for isosceles triangles with $\ks=\alpha\kl$ and $\alpha=2$, 3, or~4. For squeezed triangles (which we define as having $\alpha>4$), the formula matches the measurements up to much higher wavenumbers for $z\leq 1$, while for higher redshifts the performance is again roughly the same as the equilateral case. For brevity, in Fig.~\ref{fig:allsims-fitfunc-squeezed-z0} we only show comparisons for squeezed triangles at $z=0$, since it is for these triangles that the behavior is most different from the equilateral comparisons in Figs.~\ref{fig:tng300-fitfunc} to~\ref{fig:othersims-fitfunc-z2}. (Squeezed triangles at higher redshifts are discussed in Sec.~\ref{sec:responses} and shown in Fig.~\ref{fig:response}.)

We can understand these trends by writing the ratio of Eq.~\eqref{eq:bsc} for bispectra of fields $\delta_X$ and $\delta_{\rm DMO}$, and for isosceles triangles where two sides (denoted by $\vk_{{\rm S},1}$ and $\vk_{{\rm S},2}$) have length $\ks$ and the third has length $\kl$:
\begin{widetext}
\beq
\frac{B_X(\ks,\ks,\kl)}{B_{\rm DMO}(\ks,\ks,\kl)} = \frac{P_X(\ks)}{P_{\rm DMO}(\ks)}
	\lb \frac{F_2^{\rm (eff)}(\vk_{{\rm S},1},\vk_{{\rm S},2}) P_X(\ks)
		+ 2F_2^{\rm (eff)}(\vk_{{\rm S},1},\vk_{{\rm L}}) P_X(\kl)}
	{F_2^{\rm (eff)}(\vk_{{\rm S},1},\vk_{{\rm S},2}) P_{\rm DMO}(\ks)
		+ 2F_2^{\rm (eff)}(\vk_{{\rm S},1},\vk_{{\rm L}}) P_{\rm DMO}(\kl)} \rb\ ,
	\label{eq:fitfunc-isos}
\eeq
\end{widetext}
\noindent where we have used the fact that $F_2^{\rm (eff)}(\vk_{{\rm S},1},\vk_{{\rm L}})=F_2^{\rm (eff)}(\vk_{{\rm S},2},\vk_{{\rm L}})$. For the configurations we consider, the $P(k_{\rm L})$ term dominates over the $P(k_{\rm S})$ term in the numerator and denominator in square brackets: the power spectra are very red, scaling like $k^{n}$ with $n\leq-1$ for $k>0.1\invMpc$ , while the ratio of the respective kernels ($2F_2^{\rm (eff)}(\vk_{{\rm S},1},\vk_{{\rm L}})/F_2^{\rm (eff)}(\vk_{{\rm S},1},\vk_{{\rm S},2})$) is order unity\footnote{This behavior arises from the modified form of $F_2^{\rm (eff)}$ compared to the standard perturbation theory kernel $F_2$, for which $F_2(\vk_{{\rm S},1},\vk_{{\rm L}})/F_2(\vk_{{\rm S},1},\vk_{{\rm S},2}) \propto (\ks/\kl)^2$ for $\kl \ll \ks$.}. As long as this is true, the actual values of $F_2^{\rm (eff)}$ will mostly cancel in the ratio, making our results insensitive to whether the original fitted form from~\cite{Scoccimarro:2000ee}, the update from \cite{GilMarin:2011ik}, or even the standard perturbation theory kernel, are used. (We have verified this numerically.)

Since $P_X/P_{\rm DMO}$ is a slowly-varying function of $k$, Eq.~\eqref{eq:fitfunc-isos} will behave similarly to its equilateral counterpart when $\ks=\alpha\kl$, and therefore the formula performs essentially the same as for equilateral triangles when compared with our measurements. 
The ratio $P_X(\kl)/P_{\rm DMO}(\kl)$ is close to unity if~$\kl$ is sufficiently in the linear regime, so for squeezed triangles with $\kl$ fixed, Eq.~\eqref{eq:fitfunc-isos} will reduce to $P_X(\ks)/P_{\rm DMO}(\ks)$ to a good approximation. Our measurements of squeezed triangles also track this power spectrum ratio very closely, and therefore in this limit the fitting formula provides a closer match to the measurements. Further intuition for this result will be provided in Sec.~\ref{sec:responses}.

This style of fitting function can be extended to cross-correlations between different particle species. Eq.~\eqref{eq:bsc} amounts to the following assumption: the non-Gaussianity that gives rise to the bispectrum is solely specified by a second-order relationship between the non-Gaussian overdensity $\delta^{(X)}$ for species $X$ and a Gaussian field $\delta_{\rm G}^{(X)}$ with two-point statistics given by the nonlinear auto and cross spectra for $X$:
\begin{align*}
\delta^{(X)}(\vk) &= \delta_{\rm G}^{(X)}(\vk) \\
&\quad+ \int \frac{d^3\vq}{(2\pi)^3} F_2^{({\rm eff},X)}(\vq,\vk-\vq) 
	 \delta_{\rm G}^{(X)}(\vq)  \delta_{\rm G}^{(X)}(\vk-\vq)\ .
	 \numberthis
\end{align*}
Under this assumption, the most general cross bispectrum can be written as
\begin{align*}
B_{XYZ}(k_1,k_2,k_3) &= 2F_2^{({\rm eff},Z)}(\vk_1,\vk_2) P_{XZ}(k_1) P_{YZ}(k_2) \\
&\quad + \text{2 perms}\ .
	\numberthis
\end{align*}
When comparing this formula to measurements of the cross bispectrum between baryons and dark matter (as a ratio to the DMO auto bispectrum), we find agreement up to a maximum wavenumber between those for the baryon and dark matter auto bispectra. This follows from our previous observation that the corresponding overdensity fields are highly correlated at the scales of interest (see Fig.~\ref{fig:rbc}).

\pagebreak
\subsection{Comparison to predictions from response functions}
\label{sec:responses}

Using so-called power spectrum response functions, we can establish a point of comparison between analytical predictions and the simulation measurements of squeezed triangle configurations that is valid deep in the nonlinear regime of structure formation. These power spectrum response functions describe the ``response" of the small-scale nonlinear matter power spectrum to the presence of long-wavelength linear perturbations; they can be measured efficiently in the nonlinear regime of structure formation using separate universe simulations \citep{Li:2014sga, Wagner:2014aka, baldauf/etal:2015, andreas, Barreira:2019ckp}. From a perturbation theory point of view \citep{Barreira:2017sqa, Barreira:2017kxd}, these response functions describe squeezed mode-coupling terms, i.e., the coupling between linear long-wavelength modes with two nonlinear short-wavelength modes. 

Specifically, in the response approach, the squeezed matter bispectrum is given by
\begin{align*}
\label{eq:sqB_resp}
B(\vk_{{\rm S},1}, \vk_{{\rm S},2},\vk_L) &= \left[R_1(k_{{\rm S},1}) + \left(\mu_{\vk_{{\rm S},1}\vk_{{\rm L}}}^2-\frac{1}{3}\right)R_K(k_{{\rm S},1})\right]  \\
\numberthis
& \times P(k_{{\rm S},1})P(k_{{\rm L}}),
\end{align*}
where $\mu_{\vk_{{\rm S},1}\vk_{{\rm L}}}$ is the cosine of the angle between $\vk_{{\rm S},1}$ and $\vk_{\rm L}$, and $R_1$ and $R_K$ are the first-order responses to an isotropic overdensity and tidal field, respectively. Averaging over the angle of the bispectrum modes cancels out the $R_K$ term (for sufficiently large wavenumber bins) to yield
\begin{align*}
\label{eq:sqB_resp_aa}
\numberthis
B(\ks, \ks, \kl) = R_1(\ks) P(\ks)P(\kl),
\end{align*}
where we have specified to the case of squeezed isosceles triangles. Formally, one should also perform averages over momenta magnitude bins; we have explicitly checked however that this has a negligible impact on our results, which assume the integrands are constant within each bin.  If $\kl$ describes sufficiently large scales that are not affected by baryons, then the total matter bispectrum hydro/DMO ratio reads as
\begin{align*}
\label{eq:sqB_resp_ratio}
\numberthis
\frac{B_{\rm hydro}(\ks, \ks, \kl)}{B_{\rm DMO}(\ks, \ks, \kl)} = \frac{R_{1, \rm hydro}(\ks)}{R_{1, \rm DMO}(\ks) } \frac{P_{\rm hydro}(\ks)}{P_{\rm DMO}(\ks)};
\end{align*}
i.e., it is given by the product of the impact that baryons have on the small-scale power spectrum $P(\ks)$ and on its first-order isotropic response $R_1(\ks)$. The latter can be further decomposed into
\begin{align*}
\label{eq:R1dec}
\numberthis
R_1(k) = 1 - \frac{1}{3}\frac{{\rm dln}P(k)}{{\rm dln}k} + G_1(k),
\end{align*}
where $G_1(k)$ is the so-called growth-only power spectrum response function; this is the piece that strictly requires separate universe simulations to be measured, with the remainder of the response function being fully specified by the nonlinear power spectrum. 

\begin{figure*}
\includegraphics[width=0.985\textwidth]{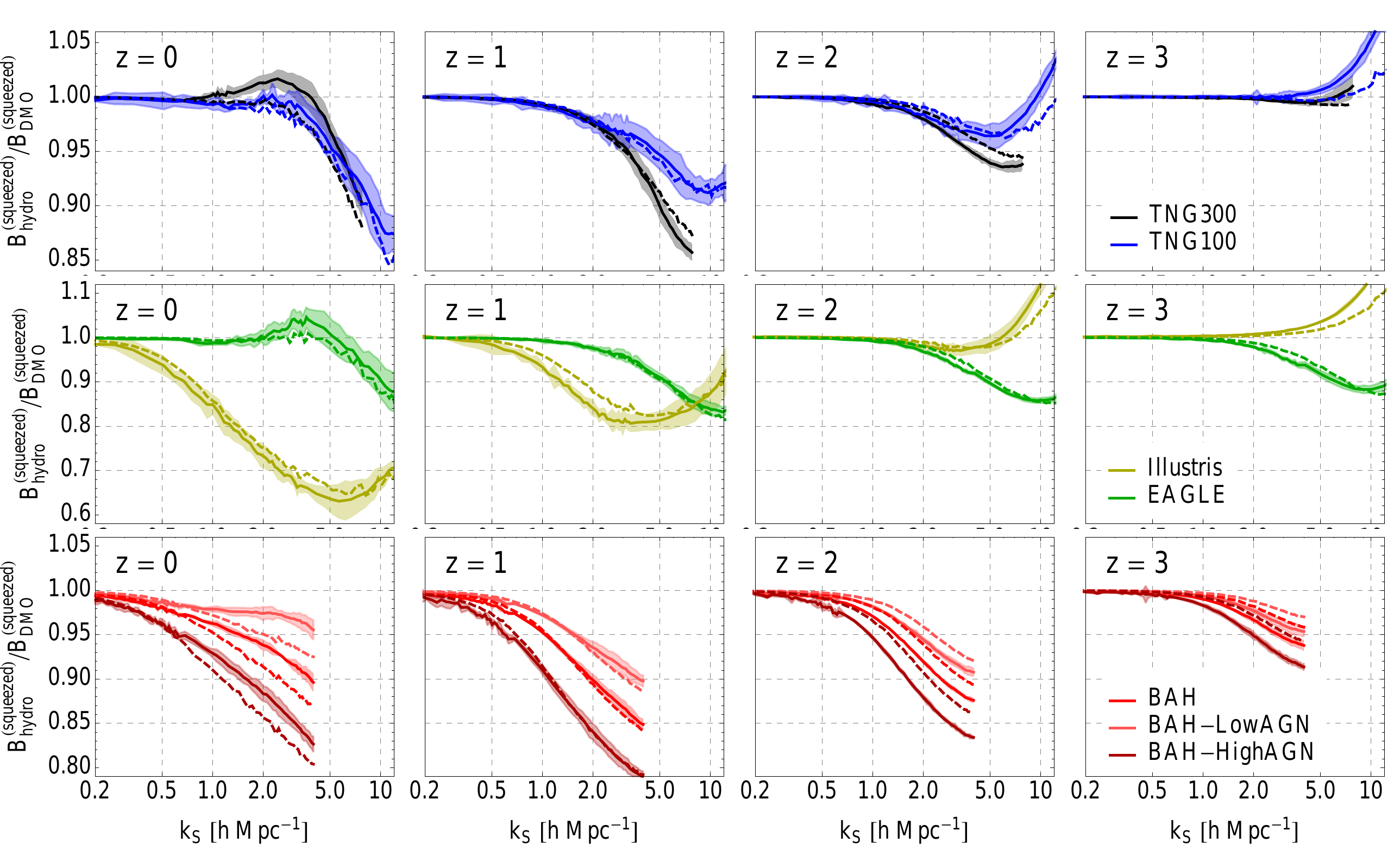}
\caption{\label{fig:response}
Comparison of predictions for total matter based on the response function formalism (dashed lines) to measurements of the squeezed matter bispectrum (solid lines) in various simulations, with the same values of $\kl$ as Fig.~\ref{fig:allsims-fitfunc-squeezed-z0}. Across all redshifts, the response result reproduces well the ratios measured in the simulations on scales as nonlinear as $k = 5\ \invMpc$. At low redshift $z<1$, the agreement for the higher resolution simulations Illustris, Eagle and TNG100 extends even to  $k = 10\ \invMpc$. The agreement with the Bahamas set of simulations is less satisfactory, with a few $\%$ differences occurring already on scales $k > 0.4\ \invMpc$.}
\end{figure*}

\cite{Barreira:2019ckp} measured $G_1$ using separate universe simulations with the IllustrisTNG model and found it to be effectively unaffected by baryonic effects at the resolution of the TNG300 box with eight times less particles than the TNG300 resolution considered here. In the comparisons we discuss next, we assume that $G_1$ remains unnafected by baryonic processes for all of the hydrodynamical implementations considered.\footnote{Specifically, we use the measurements of $G_1$ presented in \cite{Wagner:2014aka} using 16 realizations of $G_1$ from gravity-only separate universe simulations.} This way, all of the baryonic dependence of the squeezed bispectrum becomes uniquely specified by the corresponding dependence of the nonlinear matter power spectrum in Eqs.~(\ref{eq:sqB_resp_aa}) and (\ref{eq:R1dec}). In order to reduce the numerical noise in the calculation of the logarithmic derivative of the power spectrum, we evaluate the spectra as 
\begin{align*}
\label{eq:Pnl_resp}
\numberthis
P(\ks) = P_{\rm Halofit}(\ks) \frac{P_{\rm hydro}(\ks)}{P_{\rm DMO}(\ks)},
\end{align*}
where $P_{\rm Halofit}(k)$ is the {\sc Halofit} prediction \citep{halofit, halofit-update} and the power spectra ratio is the impact of baryons on the nonlinear power spectrum of each hydrodynamical simulation we consider.

Fig.~\ref{fig:response} compares the predictions of Eq.~(\ref{eq:sqB_resp_ratio}) with the corresponding simulation measurements for squeezed triangles. The level of agreement is overall satisfactory, although it varies from one simulation to the other. Specifically, the response prediction remains almost always within the EAGLE simulation's bispectrum errorbars, for all redshifts and scales shown ($z \leq 3, k \lesssim 10 \invMpc$). The agreement with Illustris, TNG300 and TNG100 is also very good, although at high redshift and small scales some differences arise; for Illustris there is also an offset at $z=1$ at $k \in [0.5, 5]\ \invMpc$ and the response result also does not reproduce the bump seen for TNG300 at $z=0$ around $k \approx 3\ \invMpc$. The BAHAMAS set is that for which the agreement is the least satisfactory: for the three cases, the response result overpredicts the amount of baryonic suppression at $z=0$ and underpredicts it at $z>2$; at $z=1$, the agreement is very good, although likely coincidental. We have also verified that the comparison between the response and BAHAMAS results remains unchanged if one adopts a different linear regime value of $\kl$.

We have specifically examined the baryonic impact on the response function ${R_{1, \rm hydro}(\ks)}/{R_{1, \rm DMO}(\ks) }$ in Eq.~(\ref{eq:sqB_resp_ratio}), which is encoded in the logarithmic derivative of the power spectrum in Eq.~(\ref{eq:R1dec}). This dependence is appreciably less pronounced than that of the power spectrum ratio (not shown), effectively rendering 
\begin{align*}
\label{eq:sqB_resp_ratio_2}
\numberthis
\frac{B_{\rm hydro}(\ks, \ks, \kl)}{B_{\rm DMO}(\ks, \ks, \kl)} = \frac{R_{1, \rm hydro}(\ks)}{R_{1, \rm DMO}(\ks) } \frac{P_{\rm hydro}(\ks)}{P_{\rm DMO}(\ks)} \approx \frac{P_{\rm hydro}(\ks)}{P_{\rm DMO}(\ks)}
\end{align*}
a good approximation. This explains also the good agreement observed in the previous subsection between the simulations and the formulae of \cite{Scoccimarro:2000ee, GilMarin:2011ik}, obtained under the assumption that the fitted effective mode-coupling kernel $F_2^{\rm eff}(\vk_{S,1}, \vk_{L})$ is not modified by baryonic effects.

Here, we do not pursue a detailed explanation for the slightly varying degree of agreement between the theoretical prediction and the various simulations in Fig.~\ref{fig:response}, specially for the three BAHAMAS cases. A possibility could be that, although \cite{Barreira:2019ckp} showed that $G_1$ in Eq.~(\ref{eq:R1dec}) is not affected by baryonic processes in TNG300, that may not necessarily be the case for other feedback implementations. As a test, we have checked that $1\%$- to $3\%$-level changes to the amplitude of $G_1$ on nonlinear scales can bring the theoretical prediction into much closer agreement with the BAHAMAS simulation results. This leaves open the possibility of $G_1$ displaying some dependence on the baryonic processes of the BAHAMAS set of simulations, which would be interesting to investigate in the future. The growth-only response $G_1$ is also expected to depend weakly on cosmological parameters, which may contribute to the small differences observed (note however that Illustris and BAHAMAS share the same cosmology, but the theoretical prediction is in good agreement with Illustris).

%--------------------------------------------------------------------------------------
% CONCLUSIONS
%--------------------------------------------------------------------------------------
\section{Conclusions}
\label{sec:conclusions}

In this paper, we have presented new measurements of total matter, dark matter, and baryon density bispectra from a selection of recent hydrodynamical cosmological simulations: IllustrisTNG, Illustris, EAGLE and BAHAMAS. These measurements are shown in detail in Figs.~\ref{fig:tng300-psbs}-\ref{fig:tng100-psbs} and~\ref{fig:tng300-bsisoc}-\ref{fig:bah-hiAGN-psbs}, and summarized in Fig.~\ref{fig:summary}. By computing ratios of measurements from the hydro and DMO simulation runs, we are able to cancel out the majority of the sample variance in the individual measurements, and we have also estimated the remaining sample variance in these ratios. We have made these measurements (along with the code used to make them) publicly available (see footnotes~\ref{foot:meas} and~\ref{foot:code}), in the hope that they are useful to the community for quantifying baryonic effects on large-scale structure and developing techniques to model or mitigate these effects in future cosmological analyses.

We have identified a feature in the total matter bispectrum ratios that is absent in the corresponding power spectrum ratios: a local maximum at $k\sim 3\invMpc$ (see Fig.~\ref{fig:summary}). This feature is also present (and has been seen previously) in the dark matter power spectrum ratios. We have found that it can be interpreted using the halo model, likely as a consequence of decreased AGN feedback strength at later times, which leads to gas re-accretion that locally alters the gravitational potentials associated with these halos. The appearance of this feature indicates that the matter bispectrum (as probed through weak lensing, for example) can provide information about the behavior of baryons that is qualitatively distinct from that obtained from the matter power spectrum. Specifically, this feature is sensitive to the time evolution of feedback at $z\lesssim 1$, and thus measuring the bispectrum would provide a method to refine our understanding of this time evolution. Moreover, in cosmic shear observations, the observed two- and three-point functions are different redshift weightings of the matter power spectrum and bispectrum, which will help to further break degeneracies in the two-point function on its own.

We have also looked specifically at quasi-linear scales (which are most useful for cosmological constraints),  finding in Fig.~\ref{fig:bs-lowk} that, as with the power spectrum, the size of the impact of baryons on the matter bispectrum varies between different simulations. For IllustrisTNG and EAGLE, the impact on the bispectrum is less than 1\%, while in BAHAMAS (which \citealt{vanDaalen:2019pst} argues is more realistic, because it is calibrated on observables more directly related to the matter power spectrum), the impact can be as large as 10\% around $k\sim 1\invMpc$. This is of the same order as the impact of massive neutrinos. Meanwhile, the scale-dependence of the relative deviation between bispectra in hydro and DMO runs scales like~$k^2$ over a wavenumber range that widens with increasing redshift, as expected from perturbative descriptions of two gravitationally-interacting fluids (baryons and dark matter) with different small-scale behaviors \citep{Lewandowski:2014rca}.

Finally, we have compared these bispectrum measurements to two modelling approaches in the literature. In Figs.~\ref{fig:tng300-fitfunc}-\ref{fig:allsims-fitfunc-squeezed-z0}, we find that the fitting functions of \cite{Scoccimarro:2000ee} and \cite{GilMarin:2011ik} perform better for simulations with weaker AGN feedback, matching the measured bispectrum ratios (within sample-variance uncertainties) up to $k\sim0.25\invMpc$ for the ``high AGN" BAHAMAS run at $z=0$ but up to $k\sim 1.2\invMpc$ for EAGLE, and even farther at higher redshifts. Conveniently, this performance is insensitive to the precise form of the mode-coupling kernel~$F_2^{\rm (eff)}$ used in the fitting formula, with bispectrum ratios depending mostly on the ratios of the relevant power spectra. Agreement with predictions from the response function formalism (e.g. \cite{Barreira:2017sqa}) for squeezed bispectrum configurations is generally even better (see Fig.~\ref{fig:response}), and this formalism has the added benefit of providing intuition for this agreement. We find possible indications that the response of the small-scale power spectrum to a long-wavelength perturbation, as encoded in the growth-only linear response function $G_1(k)$, differs between BAHAMAS and the other simulations we consider, and this would be useful to look into further. 

These results raise a number of issues that would be worthwhile to pursue in future work:
\begin{itemize}
\item There exist several approaches to modelling baryonic effects on the matter power spectrum, including modified halo models~\citep{Semboloni:2012yh,Mohammed:2014mba,2015MNRAS.454.1958M,Debackere:2019cec}, ``baryonification" prescriptions that modify outputs of DMO simulations~\citep{2015JCAP...12..049S,Schneider:2018pfw}, and relations between power spectrum suppression and baryon fractions in group- or cluster-sized halos~\citep{vanDaalen:2019pst}. Many of these approaches also automatically provide predictions for the matter bispectrum, and these predictions should be compared to our measurements as a consistency test. If this test is passed, these approaches can then be used to forecast the extent to which the bispectrum could be used to break degeneracies between baryonic and other effects that are present in the power spectrum. Further joint analyses could also be considered, incorporating other probes of baryon distributions such as the thermal Sunyaev-Zel'dovich effect (e.g. \citealt{Ma:2014dea,Tanimura:2019whh}).
\item Techniques for mitigating baryonic impacts on measured lensing statistics, such as principal component analysis~\citep{2015MNRAS.454.2451E,Huang:2018wpy}, could be extended to lensing bispectra, and assessed using forecasts. Similarly, schemes for compressing the full bispectrum down to a smaller number of measurable quantities~\citep{Byun:2017fkz,Gualdi:2018pyw} could be assessed for their sensitivity to baryonic effects.
\item Our investigation of halo profiles in IllustrisTNG revealed some features that are not easily explainable by the simple picture of late-time gas re-accretion discussed in Sec.~\ref{sec:halomodel}. It would be interesting to examine cross-matched halos between hydro and DMO runs in more detail, with a focus on the halo outskirts in particular, and also at different redshifts, to elucidate the physical origin of these effects and connect them to various power spectra and bispectra.
\item In Fig.~\ref{fig:rbc}, we found that the dark matter and baryon overdensity fields are highly correlated over a wide range of scales. This could have interesting implications for predicting the dark matter distribution from baryon-only observations, or for modelling approaches that relate the clustering of each species.
\item Motivated by recent work on constraining neutrino mass using the bispectrum~\citep{Coulton:2018ebd,Chudaykin:2019ock,Hahn:2019zob}, it would be interesting to quantify degeneracies between neutrino mass and baryonic effects, and propagate these through to forecasts for upcoming experiments.
\item Jointly fitting the free parameters in two-fluid perturbation theory~\citep{Lewandowski:2014rca,Chen:2019cfu} to the power spectrum and bispectrum will provide a stringent test of the associated predictions, and could guide further work within this framework.
\end{itemize}

As the next generation of galaxy surveys come online, and future galaxy and line intensity mapping~\citep{Ansari:2018ury} surveys are planned, more precise measurements of higher-point correlations in large-scale structure will become possible, requiring a robust theoretical understanding of these measurements. Understanding baryonic effects on the matter bispectrum will be an important step towards using these measurements to address major open questions about the Universe.

%--------------------------------------------------------------------------------------
% DATA AVAILABILITY
%--------------------------------------------------------------------------------------
\section*{Data availability}

The power spectrum and bispectrum measurements used in this work are available at the link in footnote~\ref{foot:meas}, while the code used to make these measurements can be found at the link in footnote~\ref{foot:code}.

%--------------------------------------------------------------------------------------
% ACKNOWLEDGMENTS
%--------------------------------------------------------------------------------------
\section*{Acknowledgements}

We thank Anthony Challinor, Neal Dalal, Shy Genel, Jonathan Hung, Mathew Madhavacheril, Ian McCarthy,  Alex Mead, and Martin White for useful discussions. We thank the Illustris, IllustrisTNG, and EAGLE teams for making their simulation outputs publicly available, and Ian McCarthy for providing outputs from the BAHAMAS simulations. This work made use of the Sunnyvale cluster at CITA, the TIGER cluster at Princeton and the RUSTY cluster at the Flatiron Institute. We thank John Dubinski for valuable technical support. Research at the Perimeter Institute is supported in part by the Government of Canada through the Department of Innovation, Science and Economic Development Canada and by the Province of Ontario through the Ministry of Economic Development, Job Creation and Trade. The work of FVN has been supported by the Simons Foundation. AB acknowledges support from the Starting Grant (ERC-2015-STG 678652) "GrInflaGal" of the European Research Council. WRC acknowledges support from the UK Science and Technology Facilities Council (grant number ST/N000927/1).

%--------------------------------------------------------------------------------------
% REFERENCES
%--------------------------------------------------------------------------------------
\bibliographystyle{mnras_sjf}
\bibliography{references}

\appendix
%--------------------------------------------------------------------------------------
% APPENDIX: Bispectrum aliasing
%--------------------------------------------------------------------------------------
\section{Aliasing of bispectrum measurements}
\label{app:aliasing}

\begin{figure}
\includegraphics[width=\columnwidth]{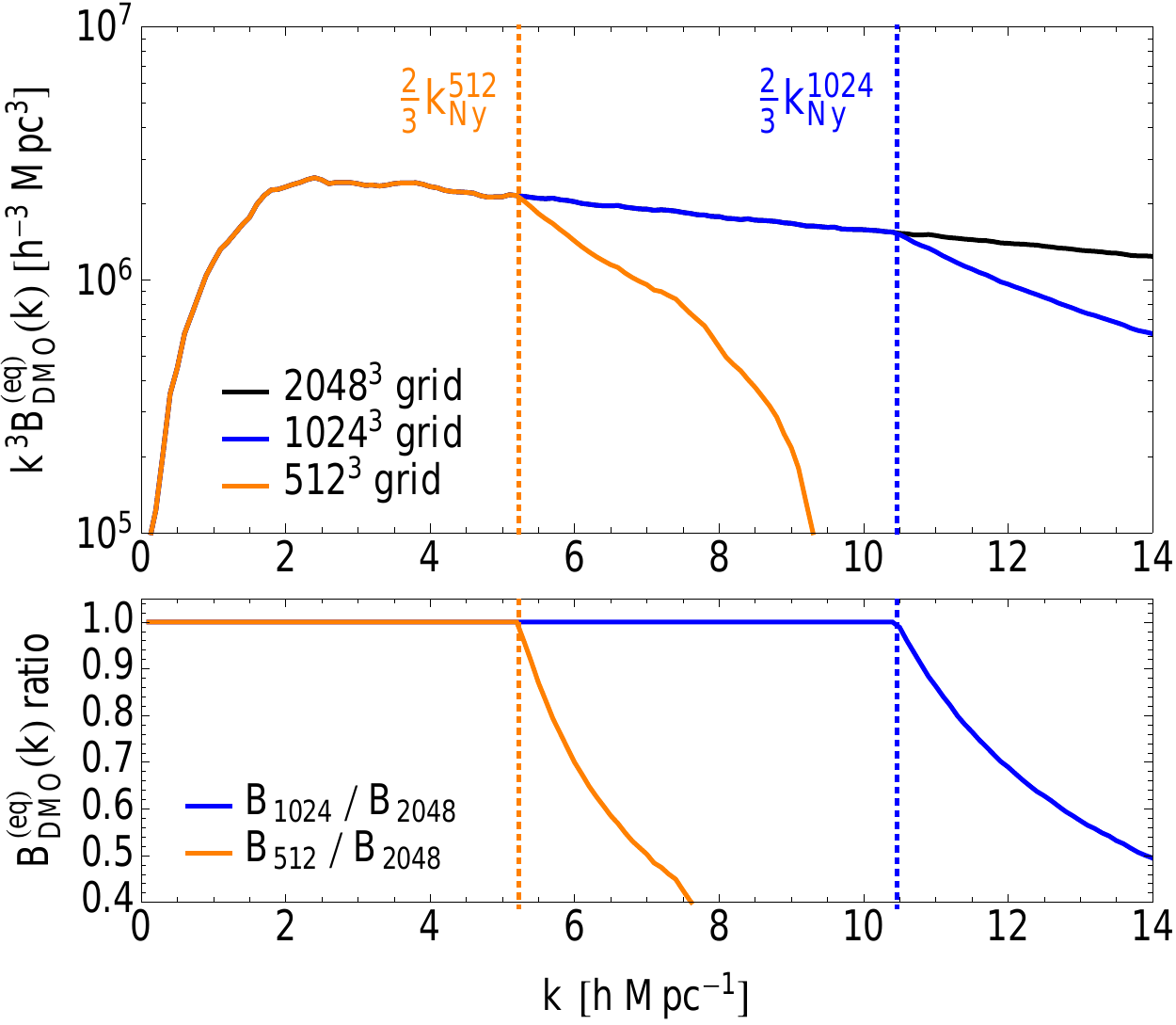}
\caption{\label{fig:b-kny}
{\it Upper panel:} The equilateral matter bispectrum measured from the DMO run of TNG300-1 using Fourier grids of resolution $2048^3$ (black line), $1024^3$ (blue line), and $512^3$ (orange line). We see a sharp change of behavior at $k=(2/3)k_{\rm Ny}$, where $k_{\rm Ny}=\pi N_{\rm grid}/L_{\rm box}$, in agreement with the expectation from~\protect\cite{Sefusatti:2015aex} that bispectrum measurements made using Eq.~\eqref{eq:best-fast} will experience aliasing at this scale. For the two lower-resolution grids, $k=(2/3)k_{\rm Ny}$ is indicated by the vertical lines. {\it Lower panel:} Ratios of measurements from the two lower-resolution grids to those from the $2048^3$ grid.
}    
\end{figure}

\cite{Sefusatti:2015aex} and~\cite{Hung:2019ygc} disagree about whether the equivalent of the Nyquist frequency for the bispectrum measured using Eq.~\eqref{eq:best-fast} is $(2/3)k_{\rm Ny}$ or $k_{\rm Ny}$, where $k_{\rm Ny}= \pi N_{\rm grid}/L_{\rm box}$ and $N_{\rm grid}$ is the number of Fourier grid cells per coordinate axis. In Fig.~\ref{fig:b-kny}, we compare the equilateral bispectrum of the DMO run of TNG300-1 measured using three different Fourier grid resolutions. We see a sharp change of behavior at wavenumbers greater than $(2/3)k_{\rm Ny}$, in agreement with~\cite{Sefusatti:2015aex}. However, this conclusion depends on the triangle configurations being considered: aliasing becomes large at $k>(2/3)k_{\rm Ny}$ for equilateral triangles, while for squeezed triangles (not shown), we find that aliasing only begins to manifest when $k_{\rm S}$ is much closer to $k_{\rm Ny}$, in agreement with ~\cite{Hung:2019ygc}. 

This configuration dependence can be understood in the following manner. The bispectrum estimator in Eq.~\eqref{eq:best-fast} involves inverse Fourier transforming to real space and then multiplying three filtered fields. This multiplication in real space can be thought of as a convolution in Fourier space. This convolution means that aliasing can occur if the filtered fields have power in modes with wavenumbers whose sum satisfies  $|\vk_1+\vk_2+\vk_3| \ge 2\pi N_{\rm grid}/L_{\rm box}$, as the product field would then have power in modes with wavenumber  $k\equiv|\vk_1+\vk_2+\vk_3|$, which would be aliased to smaller wavenumbers. In particular, if any component of the sum, e.g. $k_{1x}+k_{2x}+k_{3x}$, is equal to a multiple of $2\pi N_{\rm grid}/L_{\rm box}$, it will add signal to the bispectrum estimator, as it will be indistinguishable from the desired bispectrum, which has $k_{1x}+k_{2x}+k_{3x}=0$. The condition on the wavevector sum means that the effective Nyquist frequency for equilateral configurations is $(2/3)\pi N_{\rm grid}/L_{\rm box} = (2/3)k_{\rm Ny}$, but for highly squeezed configurations (if the first filtered map only contains large scale modes), it is close to $k_{\rm Ny}$.

This configuration dependence also explains the different conclusions of \cite{Sefusatti:2015aex} and \cite{Hung:2019ygc}. The latter considers a certain integral of the bispectrum over all configurations up to the Nyquist frequency. The integral thus includes some configurations that are affected by aliasing (equilateral-like configurations) and others that are not (squeezed like configurations) resulting in a smearing of the aliasing effects. Hence these are not strongly seen.  In contrast, when we examine specific bispectrum configurations, such as the equilateral ones shown in Fig.~\ref{fig:b-kny}, this smearing is absent and aliasing will be seen when configurations have $|\vk_1+\vk_2+\vk_3| \ge 2\pi N_{\rm grid}/L_{\rm box}$.

%--------------------------------------------------------------------------------------
% APPENDIX: Ratio outliers
%--------------------------------------------------------------------------------------
\section{Outliers in bispectrum ratio measurements}
\label{app:spikes}

We occasionally see anomalous spikes in the ratio of the measured baryon and DMO bispectra (for example, at $z=1$ and~$2$ in Fig.~\ref{fig:tng100-psbs}, and at $z=3$ in Figs.~\ref{fig:bah-psbs} to~\ref{fig:bah-hiAGN-psbs}). These spikes occur when the measured bispectra have fluctuations resulting in particularly low values (greater than one sigma away from the expected mean) at scales where the bispectrum signal to noise is relatively low (S/N $\sim 2-3$). The resulting spikes in the ratio appear significant compared to our estimated errorbars because the distribution of the ratio has heavier tails than an equivalent Gaussian distribution.

To understand why, it is helpful to consider a toy model where we model the bispectrum measurements as Gaussian random variables. In the ratio of two Gaussian random variables, if the denominator has a standard deviation that is large enough compared to the mean, the variance of the ratio is technically undefined. This is because random fluctuations in the denominator can be arbitrarily close to zero and drive the ratio to arbitrarily large values. This occurs even when the variables are highly correlated (as is the case with our bispectrum measurements). These large fluctuations are relatively rare, and are not a reflection that our sample-variance errorbars, which are an estimate of the $68\%$ confidence interval and are obtained by examining the range of the distribution of the ratio in each bin, are inaccurate. Instead, these spikes are just statistical fluctuations arising from the long tails of the distribution.

\begin{figure}
\includegraphics[width=0.5\textwidth]{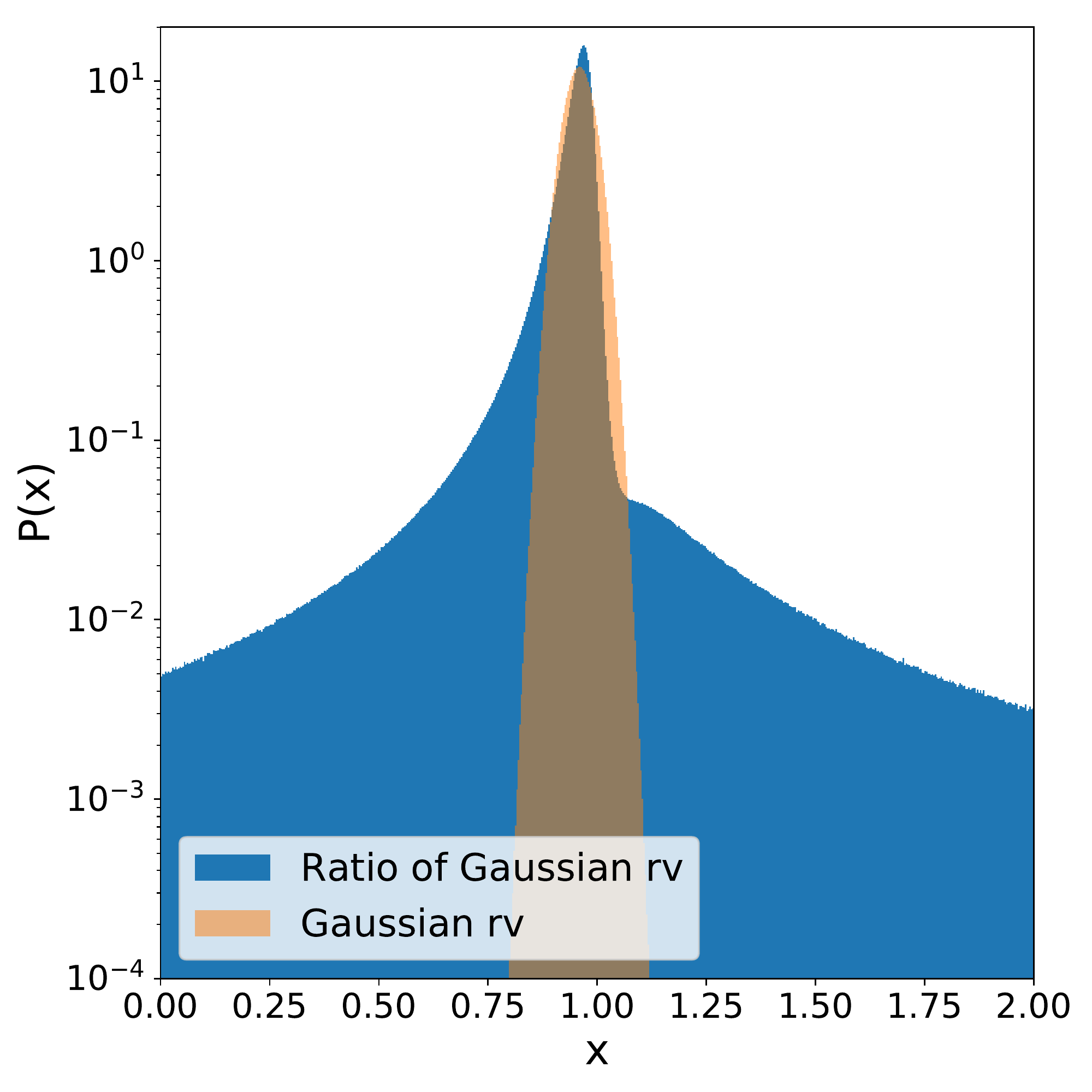}
\caption{\label{fig:grf-ratioGrf}
The probability distribution of the ratio of two Gaussian random variables (each with mean~2 and standard deviation~1, and 99.9\% correlation between them), as compared to a Gaussian distribution with the same median and 68\% range. The distribution of the ratio has significantly heavier tails than the Gaussian, increasing the probability of seeing a large outlier value in the ratio. This explains the occasional presence of anomalous spikes in our measured bispectrum ratios.
}    
\end{figure}

An example of such a heavy-tailed distribution can be seen in Fig.~\ref{fig:grf-ratioGrf}, where we plot the distribution of the ratio of two Gaussian random variables, each with mean 2 and variance 1 and which are 99.9\% correlated. For comparison, we also plot the distribution of a Gaussian random variable with the same median and 68$\%$ range. One can see that the 68\% range provides a good indication of a typical value of the ratio, but the heavy tails increase the probability of seeing a large outlier value.

%--------------------------------------------------------------------------------------
% APPENDIX: Super-sample variance
%--------------------------------------------------------------------------------------
\section{Sample variance estimation from sub-boxes}
\label{app:subboxes}

In this appendix, we justify our assertion that sample-variance uncertainties on our measurements of $P_{\rm hydro}/P_\text{DMO}$ and $B_{\rm hydro}/B_\text{DMO}$ can be estimated from the distribution of these ratios as measured from different sub-boxes of a single simulation run. Specifically, we argue that super-sample contributions to the variances of individual power spectra will mostly cancel in the variances of their ratios, and that the variances of bispectrum ratios between sub-boxes can be simply related to those from independent simulations (which do not contain super-sample effects). The same arguments also apply to the 68\% range we use for our errorbars in practice.

In general, for two random variables $X$ and $Y$ whose standard deviations satisfy $\sigma_X \ll \bar{X}$, $\sigma_Y \ll \bar{Y}$, the variance on the ratio $Z\equiv Y/X$ is given by~\citep{Villaescusa-Navarro:2018bpd}
\beq
{\rm Var}\!\lb Z \rb \approx \frac{\bar{Y}^2}{\bar{X}^2} 
	\lp \frac{{\rm Var}\!\lb X \rb}{\bar{X}^2} +  \frac{{\rm Var}\!\lb Y \rb}{\bar{Y}^2}
	- 2  \frac{{\rm Cov}\!\lb X,Y \rb}{\bar{X}\bar{Y}} \rp\ ,
	\label{eq:varZ}
\eeq
where overbars denote mean values. In our case, $X$ will denote a measurement from a DMO simulation, and $Y$ will denote the a measurement of the same quantity from the corresponding hydro run.

If the measurements are performed on sub-boxes of a larger box, the variances and covariances  will include an additive super-sample covariance (SSC) contribution related to couplings between modes contained within the sub-box and modes with longer wavelength~\citep{Hamilton:2005dx,Takada:2013bfn,Li:2014sga,Barreira:2017fjz}. Let $P_{\rm D}$ and $P_{\rm H}$ denote the ``true" (ensemble-averaged) matter power spectra in the DMO and hydro runs, measured with corresponding estimators $\hat{P}_{\rm D}$ and $\hat{P}_{\rm H}$. At wavenumbers much greater than the fundamental mode of the sub-box, the means of those estimators evaluated on the sub-box will differ by no more than a few percent from their values over the full box~\citep{Li:2014sga}. Ignoring this difference, and similarly small differences in the non-SSC (co)variances of each estimator, Eq.~\eqref{eq:varZ} implies that
\begin{align*}
&{\rm Var}\!\lb \frac{\hat{P}_{\rm H}}{\hat{P}_{\rm D}} \rb_{\rm sub}
	- {\rm Var}\!\lb \frac{\hat{P}_{\rm H}}{\hat{P}_{\rm D}} \rb_{\rm full} \\
&\; \approx \frac{\bar{P}_{\rm H}^2}{\bar{P}_{\rm D}^2}
	\lp \frac{{\rm Var}[ \hat{P}_{\rm D} ]_{\rm SSC}}{\bar{P}_{\rm D}^2}
	+ \frac{{\rm Var}[ \hat{P}_{\rm H} ]_{\rm SSC}}{\bar{P}_{\rm H}^2} 
	- 2 \frac{{\rm Cov}[ \hat{P}_{\rm D},\hat{P}_{\rm H} ]_{\rm SSC}}{\bar{P}_{\rm D}\bar{P}_{\rm H}}\rp\ .
	\numberthis
	\label{eq:pradiodiff}
\end{align*}

 Using the power spectrum response formalism~\citep{Barreira:2017sqa}, the SSC contribution to the covariance of $\hat{P}_i$ and $\hat{P}_j$ (where $i,j$ can be D or H) can be written as~\citep{Barreira:2017fjz}
\beq
{\rm Cov}\!\lb \hat{P}_i(k), \hat{P}_j(k) \rb_{\rm SSC} = \sigma_{\rm b}^2 
	R_1^{i}(k) R_1^{j}(k) \bar{P}_{i}(k) \bar{P}_{j}(k)\ ,
	\label{eq:covPAPB}
\eeq
where $\sigma_{\rm b}^2$ is the variance of the linear density field within the sub-box (as defined e.g.\ in~\citealt{Li:2014sga}), and the $R_1^i$ functions capture the response of the local power spectrum of field $i$ to a long-wavelength isotropic overdensity.\footnote{In general, the local anisotropic power  spectrum also responds to a long-wavelength tidal field, but we work with angle-averaged local power spectra, which have no such response~\citep{Barreira:2017fjz}.} These response functions are given by
\beq
R_1^i(k) = 1 + G_1^i(k) - \frac{1}{3} \frac{\d\log P_i(k)}{\d\log k}\ ,
\label{eq:R1}
\eeq
where $G_1^i(k)$ is the isotropic growth-only response function, which can be measured from separate-universe simulations~\citep{Li:2014sga,Wagner:2014aka,Barreira:2019ckp}. Combining Eqs.~\eqref{eq:pradiodiff} to~\eqref{eq:R1}, we find that
\begin{align*}
&{\rm Var}\!\lb \frac{\hat{P}_{\rm H}}{\hat{P}_{\rm D}} \rb_{\rm sub}
	- {\rm Var}\!\lb \frac{\hat{P}_{\rm H}}{\hat{P}_{\rm D}} \rb_{\rm full} \\
&\quad \approx \frac{\bar{P}_{\rm H}^2}{\bar{P}_{\rm D}^2} \sigma_{\rm b}^2
	\lp [R_1^{\rm D}]^2 + [R_1^{\rm H}]^2 -2R_1^{\rm D} R_1^{\rm H} \rp \\
&\quad = \frac{\bar{P}_{\rm H}^2}{\bar{P}_{\rm D}^2} \sigma_{\rm b}^2
	\lp G_1^{\rm D} - G_1^{\rm H} 
	- \frac{1}{3} \frac{\d\log P_{\rm D}(k)}{\d\log k} + \frac{1}{3} \frac{\d\log P_{\rm H}(k)}{\d\log k} \rp^2\ .
	\numberthis
	\label{eq:pratiodiff2}
\end{align*}
\cite{Barreira:2019ckp} measured $G_1^{\rm D}(k)$ and $G_1^{\rm H}(k)$ in hydro and DMO separate-universe simulations with the same resolution and subgrid models as TNG300-2 and -3, finding that baryonic effects have a negligible impact on these responses. Furthermore, \cite{Barreira:2019ckp} found only a small ($\sim$10\%) relative change in the logarithmic power spectrum derivatives in the hydro and DMO runs. Therefore, to a good approximation, Eq.~\eqref{eq:pratiodiff2} evaluates to zero, showing that super-sample covariance has a small impact on estimates of ${\rm Var}[\hat{P}_{\rm H}/\hat{P}_{\rm D}]$. The response functions and power spectrum tilts of dark matter and baryons within the hydro simulations also match those of the total matter quite closely, so we draw the same conclusion for them as well.

The same argument will not necessarily hold for the SSC contribution to the bispectrum covariance, since the analog of Eq.~\eqref{eq:covPAPB} for the bispectrum is not simply proportional to the mean bispectra; for an explicit demonstration of this at lowest order in perturbation theory, see~\cite{Chan:2017fiv}. Instead of pursuing further theoretical investigations of SSC for the bispectrum\footnote{For related studies, see \cite{Kayo:2012nm,Chan:2017fiv,Rizzato:2018whp,Barreira:2019icq}.}, we will instead examine simulations to measure its impact.

\begin{figure}
\includegraphics[width=\columnwidth]{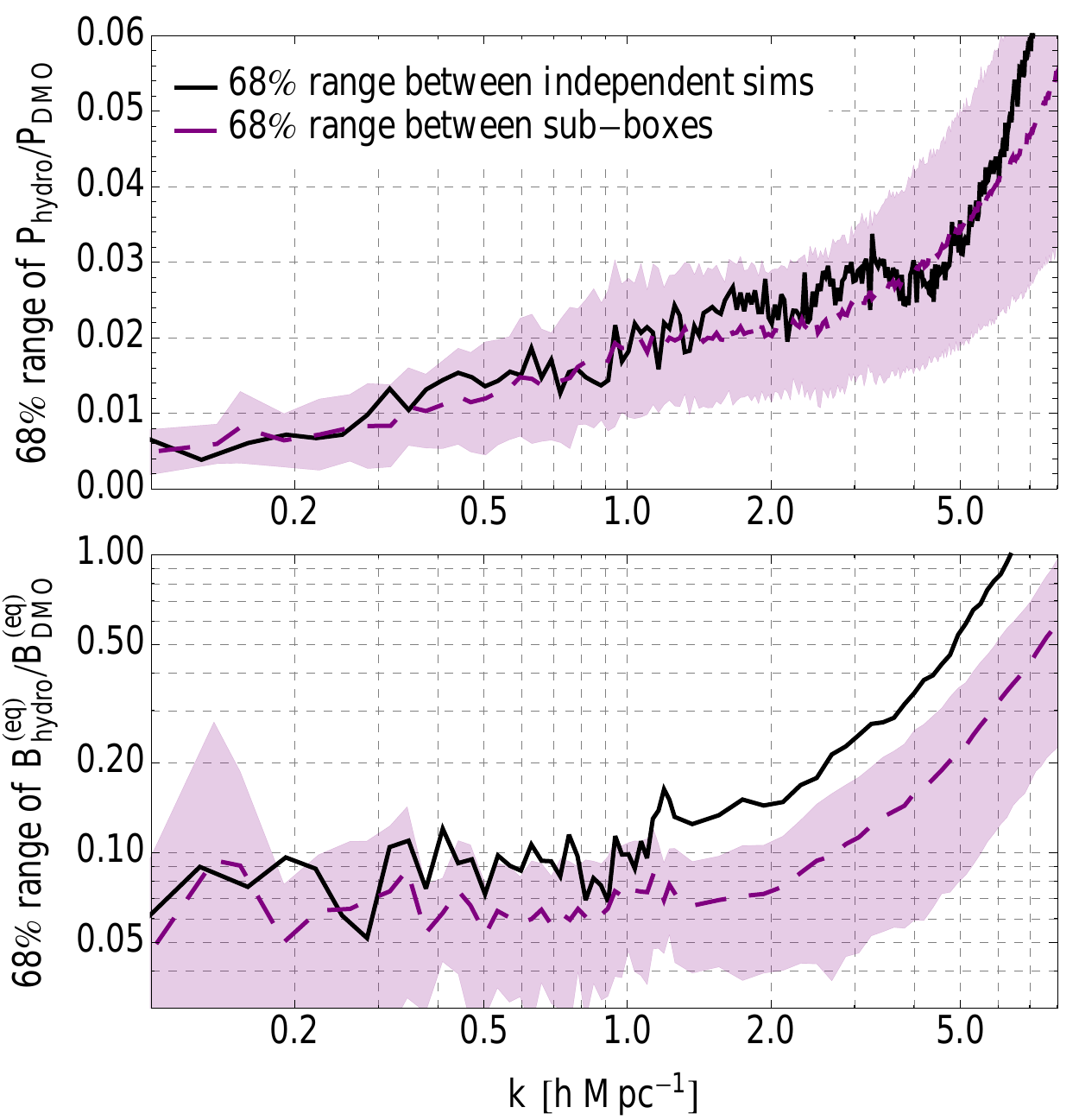}
\caption{\label{fig:subbox_var}
{\it Upper panel:} The black line shows, as a function of $k$, the 68\% range of the matter power spectrum ratio $P_{\rm hydro}/P_{\rm DMO}$ from measurements on 30 independent hydro and DMO simulation pairs with identical initial conditions (the ``H200" ensemble originally presented in~\citealt{Villaescusa-Navarro:2018bpd}). The purple dashed line and shaded region show the mean and standard deviation of 30 separate estimates of the 68\% range of this ratio, each measured from 8 sub-boxes of a single simulation pair and scaled by $1/\sqrt{8}$. The black line falls within the purple region, and therefore the variation between independent simulations is consistent with a typical result for the variation between sub-boxes of a single simulation. {\it Lower panel:} The same comparison, but for the equilateral matter bispectrum. The sub-box variations are roughly a factor of 2 lower than the independent-box variations, so we multiply the former by a factor of 2 when using them as estimates of the sample variance of our bispectrum measurements.
}    
\end{figure}

To verify our claim about the power spectrum and investigate the bispectrum ratio variances, we have made use of the ``H200" simulation set originally presented in~\cite{Villaescusa-Navarro:2018bpd}. 
This is a set of 30 hydrodynamical simulations with independent initial conditions, performed with $L_{\rm box}=200h^{-1}\,{\rm Mpc}$ and similar resolution and galaxy formation modelling to TNG300-3, and a matching set of DMO simulations with the same initial conditions. We have measured the power spectrum and bispectrum from each run, and computed the 68\% range of the hydro/DMO ratio between the different runs. For each run, we have also made the same measurements on 8 sub-boxes, computed the same range between the sub-boxes of a given run (scaling the result by $1/\sqrt{8}$ to account for the sub-box volume fraction), and computed the mean and standard deviation (over the 30 runs) of these ranges. This informs us about the distribution of ranges to be expected when using sub-boxes of a single run, and we can then check whether the range of the full independent runs is consistent with this distribution. (From our argument above, we expect consistency for the power spectrum.)

Fig.~\ref{fig:subbox_var} shows the results of this check. For the hydro/DMO ratio of the matter power spectrum (upper panel), the range between the independent simulations (black lines) is within one standard deviation of the mean of all sub-box--estimated ranges (purple regions). This comparison, along with the arguments presented above, motivates our use of the sub-box ranges as sample-variance ``errorbars" on our measurements of the ratios. For the equilateral bispectrum (lower panel), the result from independent simulations is roughly a factor of 2 higher than the typical result from sub-boxes, indicating the presence of extra correlations between sub-boxes in a given simulation that act to slightly decrease the bispectrum ratio variation between sub-boxes. To compensate for this, we multiply our sub-box--estimated bispectrum ranges by a factor of 2 when using them in our main results.

\begin{figure}
\includegraphics[width=\columnwidth]{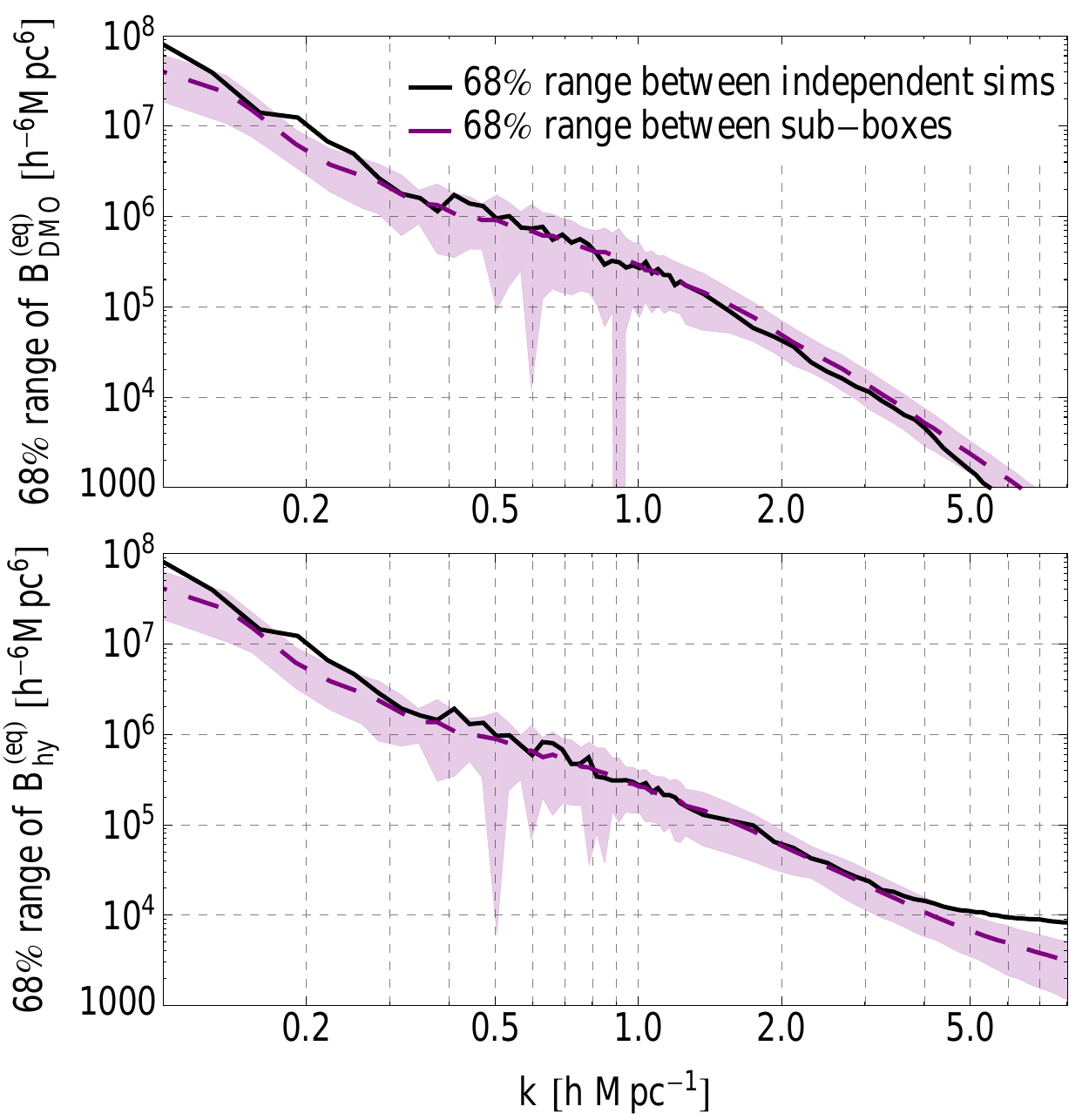}
\caption{\label{fig:bs_var}
As Fig.~\ref{fig:subbox_var}, but for equilateral matter bispectra from DMO {\em (upper panel)} and hydro {\em (lower panel)} runs separately. In both cases, the variation in bispectra measured from sub-boxes is consistent with that from independent simulations, implying that super-sample variance is not significantly affecting the sub-box measurements over the wavenumbers we show here.
}
\end{figure}

Note that this result for the bispectrum runs counter to the naive expectation that super-sample variance would increase the scatter in the measured ratios. In Fig.~\ref{fig:bs_var}, we show the analog of the lower panel of Fig.~\ref{fig:subbox_var} for the separate DMO and hydro bispectra. In both cases, we find consistency in the 68\% ranges from sub-boxes and independent-simulation, indicating that super-sample variance is not significantly affecting the sub-box measurements. This agrees with \cite{Chan:2017fiv} at quasi-linear scales ($k\lesssim 0.4\invMpc$), but no comparable study exists at deeply nonlinear scales; we leave such a study to future work. 

We have also performed this check computing standard deviations instead of 68\% ranges, and find comparable results for both the power spectrum and bispectrum.

%--------------------------------------------------------------------------------------
% APPENDIX: TNG at different resolutions
%--------------------------------------------------------------------------------------
\section{Comparison of IllustrisTNG measurements at different resolutions}
\label{app:tng-resolution}

\begin{figure*}
\includegraphics[width=0.99\textwidth]{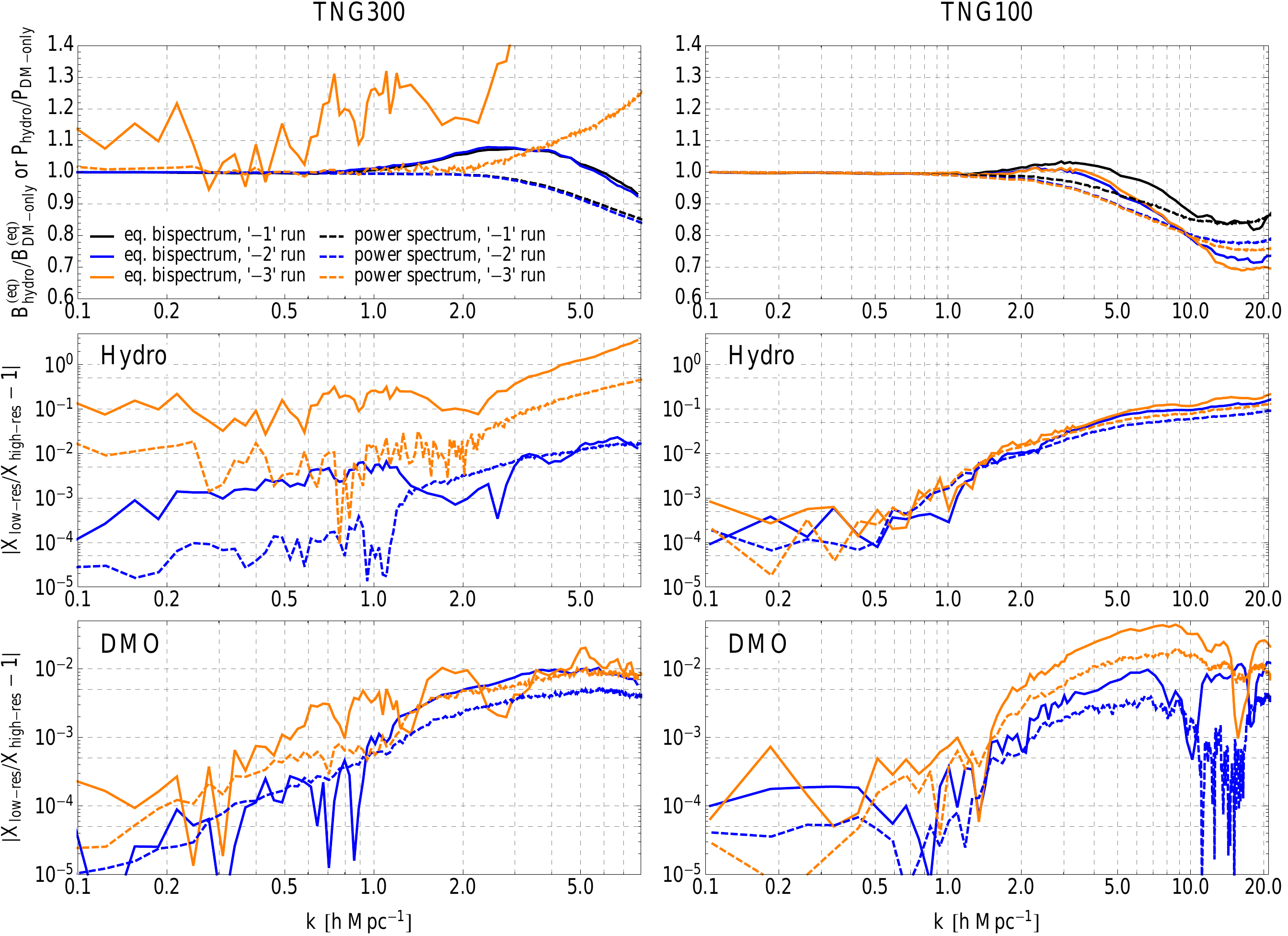}
\caption{\label{fig:tng_resolution_comparison}
Comparison of hydro/DMO power spectrum and equilateral bispectrum ratios (upper panels), along with fractional differences in separate hydro (middle panels) and DMO (bottom panels) measurements, for runs of TNG300 (left column) and 100 (right column) with different mass resolutions. TNG300-3 and -1 exhibit sizeable differences due to poor resolution of baryonic effects in the former. The TNG100 runs are slightly worse converged than TNG300-1 and -2, despite the latter having roughly eight times worse mass resolution, indicating an overall lack of convergence of large-scale clustering statistics in these simulations.
}    
\end{figure*} 

In Fig.~\ref{fig:tng_resolution_comparison}, we show the hydro/DMO ratios for the power spectrum and equilateral bispectrum from IllustrisTNG runs with different box sizes and mass resolutions (top panels), and the fractional difference between the different resolutions in the hydro (middle panels) and DMO (lower panels) runs. See Table~1 in~\cite{Springel:2017tpz} for details of the different runs. In the TNG300 runs, we find that the matter power spectra and bispectra from the hydro runs of TNG300-2 and -1 have converged to within 5\% at $k<8\invMpc$ and within 1\% at $k<4\invMpc$. In contrast, we find much poorer agreement between TNG300-3 and -1, with order-unity discrepancies in the bispectrum at $k\gtrsim 4\invMpc$. This is due to the low mass resolution of TNG300-3, which does not properly resolve baryonic effects in the lower-mass halos that contribute to these statistics and therefore exhibits much more halo-to-halo scatter in these effects than the higher-resolution simulations. Meanwhile, the DMO TNG300-3 run is in much better agreement (mostly within 1\%) with the other runs.

Interestingly, TNG100-2 and -3 also show slightly poorer agreement with 100-1 than 300-1 and -2 do with each other. TNG100-1 has roughly 8 times better mass resolution than TNG300-1, and one would naively hope for convergence of feedback effects on large-scale statistics as the mass resolution is improved, but we do not see evidence for such convergence. This provides further support for one of the conclusions of~\cite{vanDaalen:2019pst}: for reproducing large-scale observables, subgrid models within hydrodynamical simulations should be calibrated differently at different resolutions to ensure that spurious resolution-dependent effects are reduced as much as possible.

% Don't change these lines
\bsp	% typesetting comment
\label{lastpage}
\end{document}